\documentclass[manuscript,screen,nonacm]{acmart}
\AtBeginDocument{%
	}

\setcopyright{none}

\usepackage{tabularx}
\usepackage{array}
\usepackage{booktabs}
\usepackage{caption}
\usepackage{subcaption}
\usepackage{pgfplots}
\usepackage{tikz}
\usetikzlibrary{fit,shapes,arrows}
\usetikzlibrary{positioning}
\usetikzlibrary{calc,angles,quotes,arrows.meta}
\usetikzlibrary{intersections}
\usepackage{float}
\usepackage{cleveref}
\usepackage{amsthm}
\usepackage{amsmath}
\usepackage{xcolor}
\usepgfplotslibrary{fillbetween}
\pgfplotsset{compat=1.18}

\DeclareMathOperator*{\argmin}{argmin}

\newtheorem{assumption}{Assumption}[section]
\crefname{section}{Sec.}{Sections}
\crefname{figure}{Fig.}{Figures}
\crefname{table}{Tab.}{Tables}
\crefname{proposition}{Prop.}{Propositions}
\crefname{lemma}{Lemma}{Lemmas}
\crefname{theorem}{Thm.}{Theorems}
\crefname{example}{Example}{Examples}
\crefname{assumption}{Assumption}{Assumptions}
\crefname{equation}{}{}
\newcommand{\sysM}{S_{\mathrm{M}}}
\newcommand{\sysT}{S_{\mathrm{T}}}
\newcommand{\nm}{n_{\mathrm{m}}}
\newcommand{\nk}{n_{\mathrm{k}}}

\newcommand{\ntheta}{n_{\mathrm{params}}}
\newcommand{\nthetai}{n_{\mathrm{params},i}}
\newcommand{\param}{q}
\newcommand{\delay}{q_{\mathrm{d}}}
\newcommand{\area}{q_{\mathrm{A}}}
\newcommand{\radius}{q_{\mathrm{r}}}
\newcommand{\imap}{\mathcal{H}}
\newcommand{\imapA}{\hat{\mathcal{H}}}

\definecolor{mycolorGreen}{RGB}{149,189,157}%
\definecolor{mycolorGreenLight}{RGB}{213,229,216}%
\definecolor{mycolorLight}{RGB}{218,225,230}%
\definecolor{mycolorDark}{RGB}{0,45,76}%
\definecolor{mycolorRed}{RGB}{104,0,0}%
\definecolor{colT1}{RGB}{162,20,47}
\definecolor{colT2}{RGB}{217,83,25}
\definecolor{colT3}{RGB}{237,177,32}
\definecolor{colT4}{RGB}{119,172,48}
\definecolor{colT5}{RGB}{77,190,238}
\definecolor{colT6}{RGB}{0,114,189}
\definecolor{colT7}{RGB}{126,47,142}
\definecolor{colT8}{RGB}{219,169,215}
\definecolor{colT9}{RGB}{100,100,100}

\usetikzlibrary{calc}
\allowdisplaybreaks[4]

\newcommand{\sharedlegendBrake}{%
  \def\pbox{line}%
  \def\legenditems{%
    {colT1}/{line}/{T1},%
    {colT2}/{line}/{T2},%
    {colT3}/{line}/{T3},%
    {colT4}/{line}/{T4},%
    {colT5}/{line}/{T5},%
    {colT6}/{line}/{T6},%
    {colT7}/{line}/{T7},%
    {colT8}/{line}/{T8},%
    {colT9}/{line}/{T9 (after software update)}%
  }%
  \def\boxw{0.3cm}%
  \def\boxh{0.2cm}%
  \def\itemgap{0.3cm}%
  \def\labeloff{0.08cm}%
  \begin{tikzpicture}[baseline, node distance=0pt]
    \coordinate (nextbox) at (0,0);
    \foreach \col/\type/\lbl [count=\i from 0] in \legenditems {%
      \node[minimum width=\boxw, minimum height=\boxh, inner sep=0,%
            anchor=west] (box\i) at (nextbox) {};
      \ifx\type\pbox
        \fill[fill=\col] (box\i.south west) rectangle (box\i.north east);
      \else
        \node[text=\col, font=\huge] at (box\i.center) {$\ast$};
      \fi
      \node[font=\small, anchor=west, inner sep=0]
        (lbl\i) at ($(box\i.east)+(\labeloff,0)$) {\lbl};
      \coordinate (nextbox) at ($(lbl\i.east)+(\itemgap,0)$);
    }%
  \end{tikzpicture}%
}%

\begin{document}

	\title[Compositional Reachset-Conformant Modeling for Managing Uncertainty in the Development of CPS]{Compositional Reachset-Conformant Modeling for Managing Uncertainty in the Development of Cyber-Physical Systems}

	\author{Laura Lützow}
	\email{laura.luetzow@tum.de}
	\orcid{0009-0006-0745-4592}
	\affiliation{%
		  \institution{Technical University of Munich}
		  \city{Munich}
		  \country{Germany}
		}

	\author{Francesco Pio Urbano}
	\affiliation{%
		  \institution{Karlsruhe Institute of Technology}
		  \city{Karlsruhe}
		  \country{Germany}
		}

	\author{Tobias Düser}
	\affiliation{%
		  \institution{Karlsruhe Institute of Technology}
		  \city{Karlsruhe}
		  \country{Germany}
		}

	\author{Matthias Althoff}
	\affiliation{%
		  \institution{Technical University of Munich}
		  \city{Munich}
		  \country{Germany}
		}
	\email{althoff@in.tum.de}

	\renewcommand{\shortauthors}{Lützow et al.}
\begin{abstract}
The development of cyber-physical systems (CPS) proceeds through iterative design revisions, and compliance with safety requirements must be re-established after every revision, typically on the basis of a system model. Conventional models, however, predict only the most likely behavior, leading to model--plant mismatch that often surfaces only during late-stage testing, when redesign is most expensive. This article presents reachset-conformant modeling as a framework for managing uncertainty throughout the development process: instead of a single trajectory, a reachset-conformant model predicts a set of behaviors, calibrated from data to conservatively enclose the true system behavior with a statistical guarantee. Safety properties established on the model therefore transfer to the realized system. Beyond a practical introduction to reachset-conformant identification and its advantages to CPS development processes, our main theoretical contribution is a closure result: reachset conformance is preserved under serial composition, parallel composition, and feedback interconnection---and hence under any architecture assembled from these primitives. Subsystem models identified independently from local measurements can thus be substituted into the system architecture, and the interconnected model inherits a coverage guarantee from the subsystem guarantees. We complement this result with update strategies that adapt the uncertainty sets after design changes without restarting the identification workflow. Experiments on an automotive brake-system test bench demonstrate the practical benefits: a) a single model covers all operating conditions and hardware variants, b) late-stage modifications are absorbed by uncertainty set updates, c) a black-box hydraulic unit is conservatively abstracted from input--output data alone, and d) independently identified subsystem models form a conformant model by construction.
\end{abstract}
\acmArticleType{Research}



\maketitle

\section{Introduction}

\begin{figure}
    \centering
    \begin{subfigure}[t]{0.49\textwidth}
        \centering
        \resizebox{\linewidth}{!}{\newcommand{\btraj}[2]{%
\begin{scope}[shift={(-0.42,-0.41)},xscale=0.84,yscale={0.82*#1}]
\draw[#2] plot[smooth,tension=0.6] coordinates {(0.07,0.0)(0.13,0.03)(0.22,0.14)(0.31,0.47)(0.40,0.86)(0.47,1.0)(0.56,0.83)(0.66,0.67)(0.77,0.62)(0.88,0.64)(0.94,0.63)};
\end{scope}
}
\newcommand{\miniplot}[5]{%
\begin{scope}[shift={(#1,#2)}]
\begin{scope}
\clip[rounded corners=0.7pt] (-0.42,-0.41) rectangle (0.42,0.41);
\fill[white] (-0.42,-0.41) rectangle (0.42,0.41);
\fill[mycolorRed,opacity=0.85] (-0.42,0.18) rectangle (0.42,0.41);
\draw[mycolorDark!30,line width=0.3pt] (-0.355,-0.369) -- (-0.355,0.369);
\draw[mycolorDark!30,line width=0.3pt] (-0.355,-0.369) -- (0.366,-0.369);
\ifnum#5=1
\btraj{0.56}{mycolorGreen,line width=0.8pt,line cap=round}
\btraj{#3*0.42}{mycolorDark,line width=0.7pt,densely dotted,line cap=round}
\btraj{#3*0.70}{mycolorDark,line width=0.7pt,densely dotted,line cap=round}
\btraj{#3}{mycolorDark,line width=0.9pt,densely dotted,line cap=round}
\else
\btraj{#3}{mycolorGreen,line width=0.8pt,line cap=round}
\fi
\ifnum#4=1
\fi
\end{scope}
\draw[rounded corners=0.7pt,draw=mycolorDark!60,line width=0.45pt] (-0.42,-0.41) rectangle (0.42,0.41);
\end{scope}
}
\begin{tikzpicture}[font=\scriptsize,>=Latex,
stage/.style={draw=mycolorDark,semithick,rounded corners=2pt,fill=mycolorLight,text=mycolorDark,align=center,text width=0.9cm,minimum height=0.9cm,font=\scriptsize,inner sep=1pt},
release/.style={draw=mycolorDark,semithick,rounded corners=2pt,fill=mycolorDark,text=white,align=center,text width=0.9cm,minimum height=0.9cm,font=\scriptsize,inner sep=1pt},
arm/.style={->,semithick,draw=mycolorDark},
vv/.style={<->,dashed,draw=mycolorDark!35,line width=0.5pt},
loop/.style={->,line width=1.2pt,draw=mycolorRed}]
\node[stage] (ra) at (0,5.8) {Re\-quire\-ment Ana\-lysis};
\node[stage] (sd) at (0.7,4.6) {Sys\-tem De\-sign};
\node[stage] (ad) at (1.4,3.4) {Ar\-chi\-tec\-tu\-ral De\-sign};
\node[stage] (dd) at (2.1,2.2) {De\-tailed De\-sign};
\node[stage] (impl) at (2.8,1) {Im\-ple\-men\-ta\-tion};
\node[stage] (ut) at (3.5,2.2) {Unit Test\-ing};
\node[stage] (it) at (4.2,3.4) {In\-te\-gra\-tion Test\-ing};
\node[stage] (st) at (4.9,4.6) {Sys\-tem Test\-ing};
\node[stage] (at) at (5.6,5.8) {Ac\-cep\-tance Test\-ing};
\node[release] (pr) at (5.6,7.1) {Pro\-duct Re\-lease};
\draw[vv] (ra) -- (at);
\draw[vv] (sd) -- (st);
\draw[vv] (ad) -- (it);
\draw[vv] (dd) -- (ut);
\draw[arm] (ra) -- (sd);
\draw[arm] (sd) -- (ad);
\draw[arm] (ad) -- (dd);
\draw[arm] (dd) -- (impl);
\draw[arm] (impl) -- (ut);
\draw[arm] (ut) -- (it);
\draw[arm] (it) -- (st);
\draw[arm] (st) -- (at);
\miniplot{-0.9}{5.8}{0.42}{0}{0}
\miniplot{-0.2}{4.6}{0.48}{0}{0}
\miniplot{0.5}{3.4}{0.52}{0}{0}
\miniplot{1.2}{2.2}{0.56}{0}{0}
\miniplot{4.4}{2.2}{0.63}{0}{1}
\miniplot{5.1}{3.4}{0.66}{0}{1}
\miniplot{5.8}{4.6}{0.69}{0}{1}
\miniplot{6.5}{5.8}{0.92}{1}{1}
\draw[->,semithick,draw=mycolorRed] (at.north) -- (pr.south);
\node[mycolorRed] at (5.6,6.35) {\large $\mathbf{\times}$};
\draw[loop,-{>[length=6pt,width=6pt]}] (at.north) to[out=150,in=30,looseness=1] (ra.north);
\node[fill=white,draw=mycolorRed,rounded corners,inner sep=2.5pt,font=\scriptsize\bfseries,text=mycolorRed,align=center] at (2.8,7) {Unsafe behaviour detected \\$\Rightarrow$ costly redesign};
\begin{scope}[shift={(-1.2,0.9)}]
\draw[mycolorGreen,line width=0.9pt] (0,0.30) -- (0.32,0.30);
\node[anchor=west,font=\scriptsize,text=mycolorDark] at (0.40,0.30) {model prediction};
\draw[mycolorDark,line width=0.9pt,densely dotted] (0,0.075) -- (0.32,0.075);
\node[anchor=west,font=\scriptsize,text=mycolorDark] at (0.40,0.075) {observed behaviour};
\fill[mycolorRed,opacity=0.85] (0,-0.21) rectangle (0.32,-0.09);
\node[anchor=west,font=\scriptsize,text=mycolorDark] at (0.40,-0.15) {unsafe set (from specifications)};
\end{scope}
\end{tikzpicture}}
        \caption{With traditional deterministic models, model--plant mismatch can hide unsafe behaviors until late-stage testing, forcing a costly redesign loop and a delayed product release.}
        \label{fig:vmodel_traditional}
    \end{subfigure}
    \hfill
    \begin{subfigure}[t]{0.49\textwidth}
        \centering
        \resizebox{\linewidth}{!}{\input{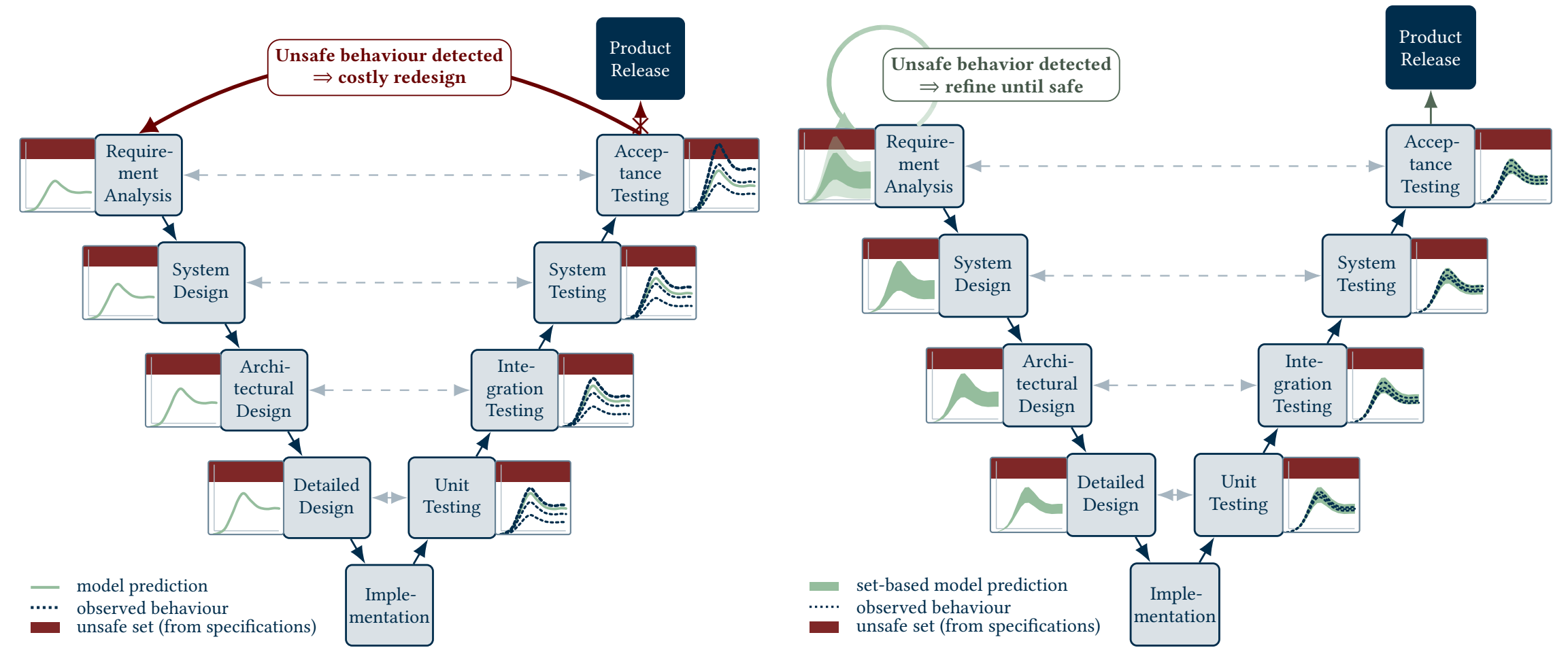}}
        \caption{With reachset-conformant models, a set of behaviors is predicted at each stage that conservatively encloses all possible system trajectories; unsafe behavior is therefore exposed during the early design stages, where redesign is cheap.}
        \label{fig:vmodel_conformant}
    \end{subfigure}
    \caption{Safety verification along the V-process, which first refines the product design from requirements analysis through system, architectural, and detailed design, and subsequently verifies the implementation through unit, integration, system, and acceptance testing.}
    \label{fig:vmodel}
\end{figure}

Cyber-physical systems (CPS) are rarely built in a single pass. Their development proceeds through a sequence of design stages---from requirements analysis through system, architectural, and detailed design down to the implementation---followed by verification stages in which units, integrated subsystems, and finally the complete product are tested against the specifications; this workflow is commonly visualized as the V-process shown in \cref{fig:vmodel}. Progress along the V is iterative: subsystem architectures, component realizations, parameters, and implementation technologies are repeatedly revised, and compliance with safety and performance requirements must be re-established after every revision. Abstract models make this feasible: derived from first principles or identified from data, they allow the system behavior to be predicted and checked against the specifications at every stage, before a concrete implementation exists.
Conventional models, however, describe most-likely rather than guaranteed behavior: exact yet tractable representations of the true dynamics rarely exist, and CPS exhibit inherent nondeterminism arising from unmodeled couplings, parameter variability, and external disturbances. The resulting model--plant mismatch may conceal unsafe behaviors until integration or acceptance testing, where the cost of correcting a defect has grown by orders of magnitude: the defect propagates back through requirements, architecture, control logic, and implementation, forcing the costly redesign loop illustrated in \cref{fig:vmodel_traditional}---precisely the cycle that early modeling is intended to prevent.

We therefore propose the automated synthesis of \emph{reachset-conformant} models~\cite{luetzow2026phd}. Where a conventional model predicts a single trajectory, a reachset-conformant model predicts a set of possible behaviors---its \emph{reachable set}---whose size reflects the model uncertainty arising from operating conditions, design choices, external disturbances, and modeling errors. The set is calibrated from measurement data such that the true system behavior is contained in it with a statistical guarantee. This containment property is termed \emph{reachset conformance}, and the computation of the evolution of the set over time is termed \emph{reachability analysis}.

This set-based view has a direct consequence for safety. Since the reachable sets of a reachset-conformant model overapproximate all possible system behaviors, safety properties verified on the model carry over to the target system~\citep{roehm2019conformance}. Unsafe behavior---indicated by an intersection of the reachable set with the unsafe region---is thereby exposed already during the early design stages, where redesign loops are inexpensive; subsequent stages primarily fix design choices and thus narrow the reachable set, so that the guarantees established early remain valid through acceptance testing without re-verification on the final hardware (\cref{fig:vmodel_conformant}). Crucially, we show in this article that reachset conformance is preserved under composition and hence compatible with modular development.

\subsection{Benefits of Reachset-Conformant Modeling}
\label{subsec:benefits}

In general, reachset-conformant modeling offers four practical benefits over conventional deterministic workflows, each demonstrated on a safety-critical brake system in \cref{sec:brakeSys}: a)~unified treatment of uncertainty due to operating conditions, design choices, and model errors, b)~systematic handling of late-stage modifications, c)~integration of black-box components, and d)~compositional modeling.

\paragraph{Unified treatment of uncertainty} Variability in environmental conditions, component realizations, and design choices forces conventional workflows to identify and validate a separate model for each relevant combination---a number that grows combinatorially---while model--plant mismatch due to structural simplifications and unmodeled dynamics deprives the verification results of formal transferability. Reachset-conformant identification instead absorbs this variability into bounded uncertainty sets embedded in the dynamics, so that a single model captures all behaviors consistent with the data and yields conservative predictions with probabilistic coverage guarantees, even for untested conditions.

\paragraph{Systematic handling of late-stage modifications} Modifications such as the substitution of a supplier component, the adjustment of a geometric dimension, or the deployment of a software update are common in industrial development, and conventionally even a minor change invalidates the affected models and necessitates a repetition of identification and validation. Reachset-conformant modeling instead accommodates such changes through an update of the uncertainty set rather than a model rebuild: a conservative inflation restores conservatism immediately, and a re-calibration on post-change measurements restores reachset conformance with statistical guarantees without restarting the identification workflow.

\paragraph{Integration of black-box components} CPS development frequently involves supplier components or legacy units whose internal dynamics are inaccessible due to restricted interfaces or proprietary algorithms, which conventionally defers system-level verification until integrated prototypes become available---precisely when design changes are most costly. Reachset-conformant identification instead constructs a conservative abstraction directly from input--output data, without knowledge of the internal model structure, so that black-box components can be incorporated into system-level models as soon as measurements are available and refined incrementally as more data is collected.

\paragraph{Compositional modeling} Since the prediction errors of deterministic subsystem models accumulate under composition in ways that are difficult to bound, conventional verification requires a monolithic system model, which is typically obtainable only after full integration; sound coverage of $N$ subsystems with $k$ realizations each then requires the analysis of $k^N$ integrated configurations. Reachset-conformant subsystem models are instead identified and validated independently from locally available measurements and subsequently composed into a system-level guarantee, reducing the effort to $N\cdot k$ subsystem analyses that proceed in parallel and commence with the first component measurements.

\subsection{Contributions and Article Structure}

This article introduces applied researchers and practitioners to the reachset-conformant modeling paradigm and demonstrates its practical benefits using an automotive brake system as case study. Complementing this, we collect practical strategies for updating reachset-conformant models after system modifications, so that a model can be adapted to a design change without restarting the identification process.
Beyond this practical perspective, our main theoretical contribution is a \emph{closure result} for reachset conformance: the property is preserved under serial composition, parallel composition, and feedback interconnection, and hence under any architecture assembled from these primitives. Since the statistical coverage guarantee composes as well, a system-level guarantee follows from independently identified subsystem models, without any system-level identification experiment.

The remainder of this article is organized as follows. \cref{sec:relWork} reviews related work on the development of safety-critical CPS and on data-driven modeling. \cref{sec:prel} introduces the reachset-conformant modeling framework, comprising safety verification via reachability analysis, reachset-conformant identification, and the novel model update strategies. \cref{subsec:composition} establishes the compositionality of reachset-conformant models. \cref{sec:brakeSys} demonstrates the practical benefits of the framework on the brake-system case study, and \cref{sec:concl} concludes the article.

\section{Related Work} \label{sec:relWork}

This section reviews conventional development processes for complex CPS and summarizes prior work on reachset-conformant identification, which forms the methodological foundation of this article.

\subsection{Uncertainty-Aware Product Development}
\label{subsec:developmentMechatronicCPS}

Uncertainty in product development has long been recognized as a central driver of iteration, redesign, and late-stage validation efforts. In robust design methodology, the objective is to reduce sensitivity to uncontrollable variation or noise factors \cite{Arvidsson2008, Acar2021, Wang2023, Beyer2007}. One line of research uses probabilistic methods to improve the likelihood that the design works by early incorporating uncertainties \cite{Calafiore2006, Du2004, Chen1996, Chen1996b}.

Similar to our work, set-based design supports uncertainty-aware product development by maintaining and progressively refining sets of feasible design alternatives rather than committing early to a single solution \cite{Toche2020} -- these approaches, however, only work for static models, and existing research still lacks holistic methodologies for integrating and managing these sets across the full product development process. When considering finite sets of static models, one can use rough set theory to determine cases in which a property will hold for sure and cases where more information is required \cite{Fan2024}. Other approaches use fuzzy sets that have a degree of membership ranging from (not in the set at all) to 1 (fully in the set) \cite{Wood1989, Antonsson1995}. Finding optimal tolerances is a special case of set-based design of static systems \cite{Morse2018} -- the goal is to find the largest possible tolerances so that the system specification of a static model is met.

Existing uncertainty-aware and set-based design methods primarily focus on maintaining, refining, and selecting sets of design alternatives under uncertain parameters and requirements, yet they typically do not model the design as a dynamical system whose behavior evolves over time. As a result, uncertainty is propagated to a set of feasible designs or performance outcomes, whereas the effect of uncertainty on the complete set of admissible system behaviors and their safety-related implications remains largely unaddressed.

\subsection{Data-Driven Modeling of CPS}

As CPS grow in complexity, purely first-principles models become increasingly difficult to derive and often fail to capture effects such as unmodeled dynamics, component variability, and environmental influences that are hard to characterize analytically. Data-driven approaches address this gap by inferring models directly from measured input-output behavior, complementing or replacing physical modeling assumptions with information extracted from experiments, and thereby enabling up-to-date models even when the underlying physics is partially unknown or subject to change.
 This data-driven paradigm is formalized by the field of system identification, which provides a well-established methodology for constructing dynamical models from measured input-output data, with a rich body of foundational textbooks and survey articles giving a comprehensive overview of the available methods \citep{ljung2002identification,isermann2010identification,keesman2011system,toth2010modeling,johan2019nonlinear,adeniran2017nonlinear,gueho2021advances,yassin2013RecentA}. Key directions include prediction-error and maximum-likelihood methods \citep{astrom1965linear,ljung1981frequency}, subspace identification \citep{ho1965effective,overschee1996subspace,gustafsson2001subspace}, frequency-domain techniques \citep{schoukens2021frequency}, nonlinear system identification techniques for block-oriented \citep{voros1999hammerstein,lin2016wiener,pearson2000block}, Volterra \citep{mirri2002volterra,glentis1999volterra}, and NARX models \citep{chen1989narx,piroddi2003narx}, and, more recently, machine-learning-based approaches \citep{pillonetto2025deepSI,chiuso2019si}.

A parallel line of research pursues bounded-error (set-membership) identification, which computes the set of parameters consistent with the measured data and an assumed noise bound \citep{walter1990survey}, using set representations such as intervals or multi-dimensional boxes \citep{kieffer2006box,polyak2005interval}, ellipsoids \citep{polyak2004ellipsoidal}, and zonotopes \citep{bravo2006bounded}. However, this requires prior knowledge of the model structure and the size of the uncertainty bounds, assumptions that are rarely easy to justify a priori: bounds chosen too conservatively inflate the resulting parameter set, while bounds chosen too tightly can leave no parameters consistent with the data.

The methods presented above are predominantly geared toward capturing average or nominal system behavior, or otherwise depend on restrictive assumptions about model structure and noise properties. This makes them a poor fit for rigorous safety verification, which requires behavioral abstractions that capture all possible system behaviors. Instead, such abstraction can be obtained using reachset-conformant identification.
Reachset conformance, initially introduced by \citet{althoff2012reachability} and formalized by \citet{roehm2016reachset}, is a formal relationship that enables the transference of safety properties: when a model provides an overapproximation of the reachable outputs of a system, it can be used as a sound substitute for verifying safety \cite[Prop.~1]{roehm2019conformance}. The key principle of reachset-conformant identification is to establish this relationship by augmenting the model with nondeterminism—referred to here as uncertainty—ensuring that all measured system trajectories remain enclosed within the reachable set of the model. Uncertainty sets with minimum size can be computed with convex programs \citep{gruber2023scalable}, while linear programming techniques can identify uncertainty sets that produce the smallest possible reachable sets \citep{Liu2023conf,luetzow2024generator}. Recent advances have extended the methodology to nonlinear models via Taylor-expansion-based linearization \citep{luetzow2026reachset} and enabled the handling of large-scale datasets via recursive schemes \citep{luetzow2025recursive}.
Beyond dynamical systems, the framework can be generalized to uncertainty quantification for arbitrary predictive models, such as neural networks trained for regression or classification tasks \citep{luetzow2025zonoconformal}, exhibiting a strong connection to conformal prediction~\cite{luetzow2026mcp,angelopoulos2026theoreticalfoundationsCP}.

\section{Reachset-Conformant Modelling}\label{sec:prel}

In this section, we introduce the fundamental concepts underlying the construction of reachset-conformant models.
Throughout this paper, we use the following notation: For a sequence of vectors $v_i \in \mathbb{R}^m$, we write $v_{0:k}:=[v_0^\top~v_1^\top~\cdots~v_{k}^\top]^\top\in \mathbb{R}^{m(k+1)}$ for their vertical concatenation. The $j$-th component of $v_i$ is written as $v_{i(j)}$. Sets are denoted by calligraphic letters. For a sequence of sets $\mathcal{V}_i \subseteq \mathbb{R}^m$, we  denote their Cartesian product by $\mathcal{V}_{0:k}:=\{v_{0:k}\mid v_i\in\mathcal{V}_i,~i=0,\dots,k\}\subseteq \mathbb{R}^{m(k+1)}$.
The image of a set-valued map $\mathcal{H}\colon \mathbb{R}^{m} \to 2^{\mathbb{R}^{n}}$ over the set $\mathcal{V} \subseteq \mathbb{R}^m$ is defined as $\mathcal{H}(\mathcal{V}) := \bigcup_{v\in\mathcal{V}} \mathcal{H}(v)$.

\subsection{Safety Verification via Reachability Analysis and Reachset Conformance} \label{subsec:RA}

Reachability analysis provides a formal mechanism for verifying safety of dynamical systems under bounded uncertainties. Consider a discrete-time nonlinear state-space model $S$:
\begin{subequations}\label{eq:generalSystem}
\begin{align}
x_{k+1} &= f(\param,x_k, u_k, w_k), \label{eq:generalSystemState}\\
y_k &= g(\param,x_k, u_k, w_k),\label{eq:generalSystemOutput}
\end{align}
\end{subequations}
where $x_k \in \mathbb{R}^{n_{\mathrm{x}}}$ denotes the state at time step $k \in \mathbb{N}$, $u_k \in \mathbb{R}^{n_{\mathrm{u}}}$ the control input, $w_k \in \mathcal{W}\subset \mathbb{R}^{n_{\mathrm{w}}}$ a bounded disturbance, and $y_k \in \mathbb{R}^{n_{\mathrm{y}}}$ the output. The dynamics and output map depend on parameters $\param \in \mathcal{Q}$.

Let $\Theta_k(S,q,x_0,u_{0:k},w_{0:k})$ denote the output at time $k$ for parameters $\param$, initial state $x_0$, input sequence $u_{0:k}$, and disturbance sequence $w_{0:k}$. The reachable output set is defined as
\begin{align*}
    \mathcal{Y}_k(S,x_0,u_{0:k})
=
\left\{
\Theta_k(S,q,x_0,u_{0:k},w_{0:k})
\;\middle|\;
w_i \in \mathcal{W},\; i=0,\dots,k,\; \param \in \mathcal{Q}
\right\}.
\end{align*}
Throughout most of the paper, $\mathcal{Q}$ is assumed to be a singleton corresponding to fixed nominal parameters, unless stated otherwise.

Safety specifications are typically expressed via avoid sets $\mathcal{Y}_{\mathrm{unsafe},k} \subseteq \mathbb{R}^{n_{\mathrm{y}}}$. The system is safe over a finite horizon $\nk$ for initial state $x_0$ and input sequence $u_{0:\nk-1}$ if
\begin{align} \label{eq:checkSafe}
    \mathcal{Y}_k(S,x_0,u_{0:k})
\cap
\mathcal{Y}_{\mathrm{unsafe},k}
= \emptyset,
\qquad \forall k = 0,\dots,\nk-1.
\end{align}
Hence, safety verification reduces to checking disjointness between reachable sets and unsafe regions.

\begin{figure}
    \centering
    \input{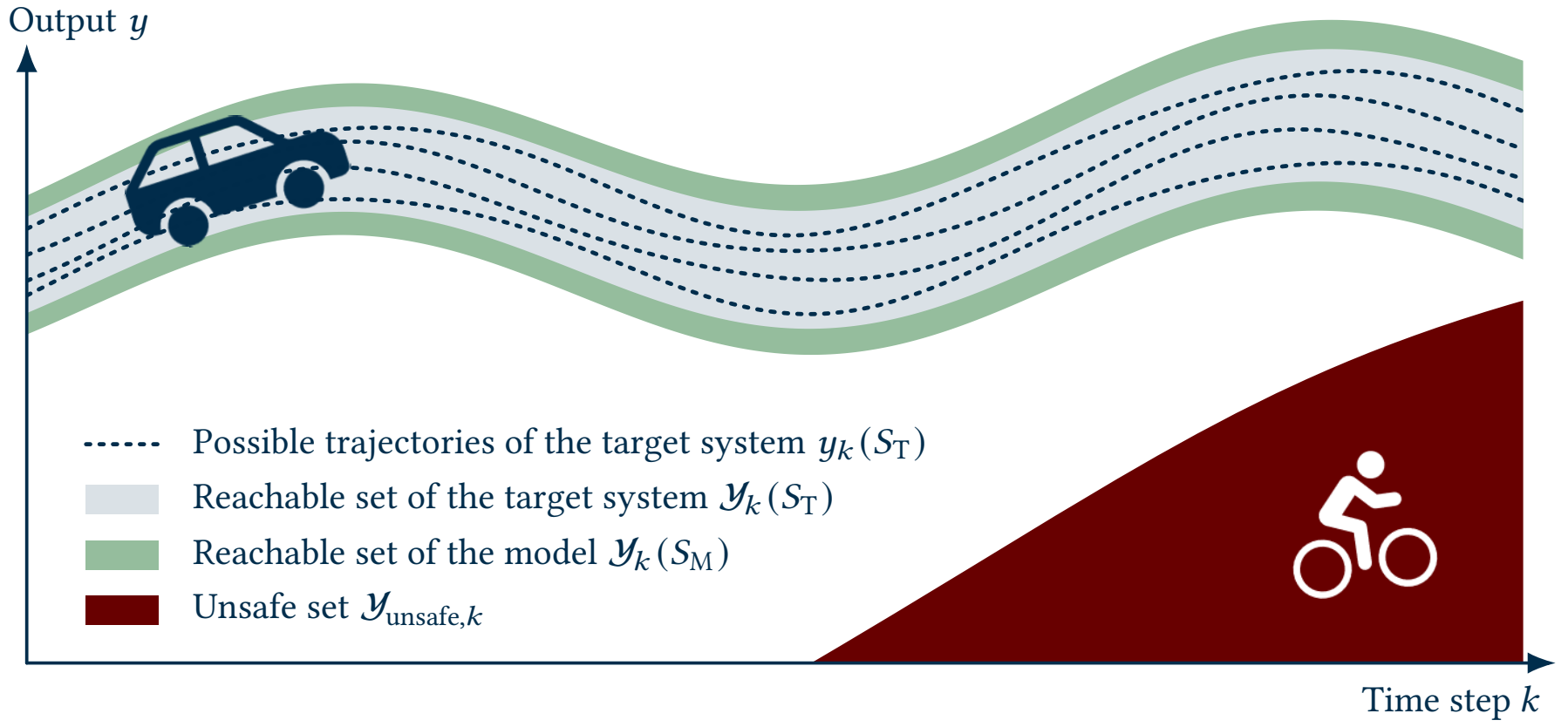}

\caption{Safety verification via reachability analysis: The target system $\sysT$ (e.g., an autonomous vehicle) is certified as safe if the reachable set $\mathcal{Y}_{k}(\sysM)$ of a reachset-conformant model $\sysM$, which overapproximates the reachable set $\mathcal{Y}_{k}(\sysT)$ of $\sysT$, does not intersect the unsafe set $\mathcal{Y}_{\mathrm{unsafe},k}$ (e.g, the reachable set of other traffic participants) for every time step $k$.}
\label{fig:verification}
\end{figure}

Since exact reachable sets can only be computed for special cases~\citep{lafferriere1999newClass}, tight over-approximations are typically computed, and
reachset-conformant models provide a systematic means of constructing these overapproximations. A model $\sysM$ is reachset-conformant to a target system $\sysT$ if, for all admissible initial states and input sequences,
\begin{align} \label{eq:reachConf}
    \forall k:\quad
\mathcal{Y}_k(\sysM,x_0,u_{0:k})
\supseteq
\mathcal{Y}_k(\sysT,x_0,u_{0:k}).
\end{align}
By checking that these overapproximations do not intersect any unsafe sets, one can prove safety as illustrated in \cref{fig:verification}.

\subsection{Reachset-Conformant Identification} \label{sec:rcModeling}

Practical methods adopt a data-driven notion of \cref{eq:reachConf}, requiring containment only for a finite collection of test cases~\cite{luetzow2026reachset}.
Let $\mathcal{Y}_k^{(m)}(\sysM):=\mathcal{Y}_k(\sysM,x_0^{(m)},u_{0:k}^{(m)})$ denote the reachable output set of a candidate model $\sysM$ under test case $m$ and let $y_{0:\nk-1}^{(m)}(\sysT)$ denote the corresponding measured outputs of the target system. Reachset-conformant identification seeks a model $\sysM$, described by model equations as in \cref{eq:generalSystem} and an uncertainty set $\mathcal{W}$, solving
\begin{subequations} \label{eq:opt}
\begin{align}
\sysM^* &= \argmin_{\sysM} ~~ \sum_{m\in\mathcal{M}_{\mathrm{eval}}} \sum_{k=0}^{\nk-1}
\mathrm{size}\big(\mathcal{Y}_k^{(m)}(\sysM)\big) \label{eq:cost}\\
\text{s.t.}\quad
& \forall k\colon \forall m\in\mathcal{M}:\;
y_k^{(m)}(\sysT) \in \mathcal{Y}_k^{(m)}(\sysM).
\label{eq:containment}
\end{align}
\end{subequations}
Constraint \eqref{eq:containment} enforces data consistency, i.e., all observed trajectories must be contained in the reachable sets. The objective in \cref{eq:cost} penalizes the size of these sets—e.g., via a proxy for their volume—evaluated over all time steps and representative test cases in $\mathcal{M}_{\mathrm{eval}}$. This discourages overly conservative models, whose excessively large reachable sets would reduce the informativeness of subsequent safety verification.

Optimally solving \cref{eq:opt} is computationally intractable. Thus, we propose the following sequential identification approach~\citep{luetzow2026reachset,luetzow2025zonoconformal}:
\begin{enumerate}
    \item \textbf{Deterministic model construction.}
    A predictive model structure is specified. This model can be derived from first principles or identified using classical system identification techniques from an additional data set. For example, consider a one-dimensional system model
    \begin{subequations}\label{eq:exSS}
    \begin{align}
        x_{k+1} &= \param_{(1)} x_k^2 + \param_{(2)} u_k^2 + \param_{(3)} x_k u_k + \param_{(4)} x_k + \param_{(5)} u_k, \\
        y_k &= \param_{(6)} x_k u_k + \param_{(7)} x_k,
    \end{align}
    \end{subequations}
    with parameters $\param \in \mathbb{R}^7$.

    \item \textbf{Uncertainty injection.}
    The deterministic model is rendered nondeterministic by introducing bounded uncertainty variables $w_k \in \mathcal{W}$. These variables capture model mismatch, external disturbances, and measurement noise. Uncertainties may be placed at physically motivated or random locations; in practice, however, it is often advantageous to also include additive uncertainty terms, as they yield broad and readily interpretable effects on the reachable set.

    Applied to \cref{eq:exSS}, this yields, for example,
    \begin{align*}
        x_{k+1} &= (\param_{(1)} + \textcolor{mycolorDark!60}{w_{k(1)}}) x_k^2
        + \param_{(2)} (u_k + \textcolor{mycolorDark!60}{w_{k(2)}})^2
        + \param_{(3)} x_k u_k
        + \param_{(4)} (x_k + \textcolor{mycolorDark!60}{w_{k(3)}})
        + \param_{(5)} u_k
        + \textcolor{mycolorDark!60}{w_{k(4)}}, \\
        y_k &= \param_{(6)} x_k u_k + \param_{(7)} x_k + \textcolor{mycolorDark!60}{w_{k(5)}}.
    \end{align*}
    Here, $w_{k(1)}$, $w_{k(2)}$ and $w_{k(3)}$ could represent structural mismatch, and $w_{k(4)}$ and $w_{k(5)}$ capture additive process and measurement uncertainty. If only an initial state estimate $x_{0,\mathrm{est}}$ is available, initial-state uncertainty is modeled as $x_0 = x_{0,\mathrm{est}} + \textcolor{mycolorDark!60}{w_{0(6)}}$.

    \item \textbf{Uncertainty-set calibration.}
    The uncertainty set $\mathcal{W}$ is calibrated such that all measured outputs satisfy the containment constraint~\eqref{eq:containment} while minimizing reachable set size. By imposing a suitable parametric structure on $\mathcal{W}$, the calibration problem can be formulated as a linear program \citep[Alg.~1]{luetzow2026reachset}, which is provided in Appendix~\ref{sec:lp}.
\end{enumerate}
The resulting nondeterministic model satisfies data-driven reachset conformance for all test cases $m \in \mathcal{M}$.
The probability that future, previously unseen trajectories are also contained in the reachable sets increases with the number of identification test cases and can be quantified using scenario theory~\cite{calafiore2006scenario}.
In particular, we obtain the following result for the coverage probability
\begin{align}\label{eq:coverageProb}
    \eta(\mathcal{M})=\mathbb{P}\{y_k^{(\tilde{m})}(\sysT)\in \mathcal{Y}_k^{(\tilde{m})}(\sysM),~k=0,\dots,\nk-1\},
\end{align}
which is the probability that the measurements $y_{0:\nk-1}^{(\tilde{m})}$ of a new test case $\tilde{m}$ lie within the reachable sets of the model.
\begin{assumption}\label{ass:coverage}
    Let the uncertainty set parameters be estimated using the linear program in \citet[Alg.~1]{luetzow2026reachset}, where the cost function is evaluated over test cases that are independent from the test cases in $\mathcal{M}$ used in the constraints. Assume that the feasibility region of the linear program has a non-empty interior and the solution exists and is unique.
\end{assumption}
\begin{proposition}[Coverage Guarantees]\label{prop:cover}
    Let all test cases in $\mathcal{M}$ and the new test case $\tilde{m}$ be exchangeable.
    Under \cref{ass:coverage}, the expected coverage probability for the new test case $\tilde{m}$ satisfies
    \begin{align}
    \mathbb{E}\{\eta(\mathcal{M})\}\geq 1-\frac{\ntheta}{\nm+1}, \label{eq:expectedCov}
    \end{align}
    where $\ntheta$ denotes the number of parameters identified over $\mathcal{M}$ and $\nm$ is the number of test cases in $\mathcal{M}$.
\end{proposition}
The proof is deferred to Appendix~\ref{app:proofs} to maintain reading fluency.

\subsection{Model Updates}
\label{subsec:modelUpdates}
\begin{figure}[t]
\centering
\resizebox{0.92\linewidth}{!}{\input{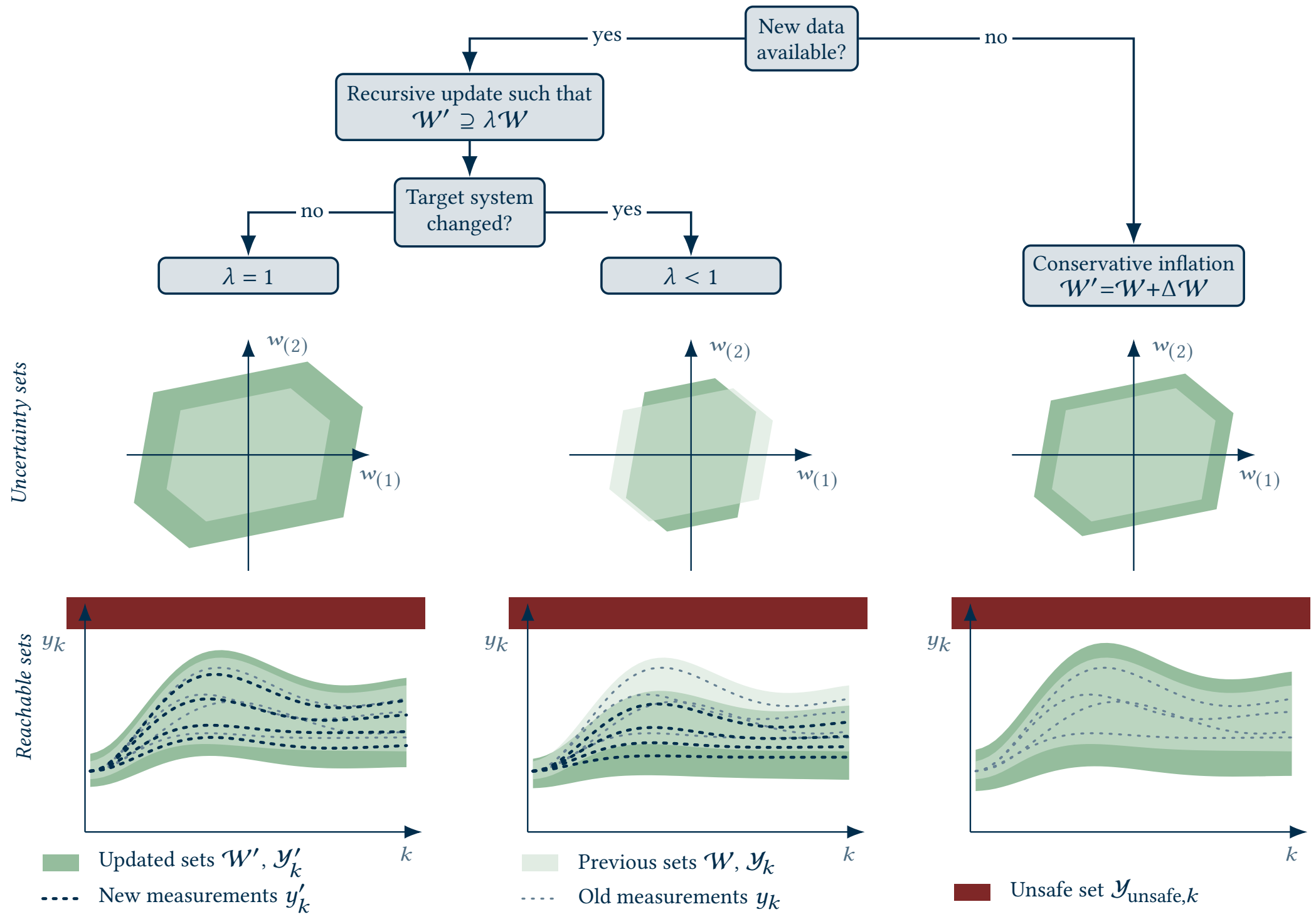}}
\caption{Mechanics of the model-update strategy. \textbf{Case~1} (new data, unchanged system): the updated reachable sets $\mathcal{Y}_{0:\nk-1}'\supseteq\mathcal{Y}_{0:\nk-1}$ cover old and new trajectories since $\mathcal{W}'\supseteq\mathcal{W}$. \textbf{Case~2} (new data, changed system): $\mathcal{Y}_{0:\nk-1}'$ might no longer cover all old trajectories but enclose all new ones. \textbf{Case~3} (no data, changed system): $\mathcal{Y}_{0:\nk-1}'\supseteq\mathcal{Y}_{0:\nk-1}$ since $\mathcal{W}'=\mathcal{W}+\Delta\mathcal{W}$ is conservatively inflated.}
\label{fig:modelUpdates}
\end{figure}

In practical development processes, models are rarely static: additional measurement data become available over time, new operating conditions are introduced, and system components are modified. Any such change may invalidate previously established safety guarantees and therefore calls for a rapid model update rather than a complete re-identification from scratch. A key strength of the reachset-conformant framework is that these updates reduce to an adaptation of the uncertainty set $\mathcal{W}$, while the deterministic model equations can remain fixed. We distinguish two situations that a practitioner typically faces: (i)~additional data becomes available, and (ii)~a structural change must be assessed before any post-change data exists. The corresponding update mechanisms are illustrated in \cref{fig:modelUpdates}.

\paragraph{Additional data} When additional data $\mathcal{M}'$ becomes available, the uncertainty set can be updated efficiently with one of the recursive methods of \cite{luetzow2025recursive}, without revisiting the original data set $\mathcal{M}$. The most efficient variant establishes reachset conformance with respect to $\mathcal{M}'$ while enforcing $\mathcal{W}'\supseteq \lambda\,\mathcal{W}$ with $\lambda\in[0,1]$, where $\mathcal{W}$ and $\mathcal{W}'$ denote the previous and updated uncertainty set, respectively. The parameter $\lambda$ encodes how strongly the practitioner trusts the old data $\mathcal{M}$ to remain representative of the current target system. Using the reachset-conformant identification approach described in Appendix~\ref{sec:lp}, the recursive updates can be done via linear programming.

If the system has \emph{not} changed (\textbf{Case~1} in \cref{fig:modelUpdates}), one sets $\lambda=1$. By enforcing $\mathcal{W}'\supseteq\mathcal{W}$, the updated reachable sets $\mathcal{Y}_{0:\nk-1}'$ must enclose the previous reachable sets $\mathcal{Y}_{0:\nk-1}$, i.e., $\mathcal{Y}_{0:\nk-1}'\supseteq \mathcal{Y}_{0:\nk-1}$, which ensures reachset conformance over the combined data set $\mathcal{M}\cup\mathcal{M}'$. As the data set grows, the probability that future trajectories lie inside the reachable sets increases according to the scenario-theoretic bound in \cref{eq:expectedCov}, so incorporating new data systematically improves robustness against unseen behavior.

If the system \emph{has} changed between the collection of $\mathcal{M}$ and $\mathcal{M}'$ (\textbf{Case~2} in \cref{fig:modelUpdates}), one chooses $\lambda<1$. This allows the model to become less conservative by relaxing conformance with respect to past test cases in $\mathcal{M}$: the updated set $\mathcal{W}'$ can expand but also slowly contract along some directions to capture the new behavior. Thus, the updated reachable sets $\mathcal{Y}_{0:\nk-1}'$ need no longer cover every old trajectory but are still guaranteed to enclose all new ones. The limiting case $\lambda=0$ discards the past entirely and is equivalent to a fresh identification over $\mathcal{M}'$, the appropriate choice for a substantially changed system. The parameter $\lambda$ thus provides the practitioner with a single, interpretable dial for trading off retained robustness against adaptation to the new regime.

\paragraph{System change without available data} Design changes---revised requirements, modified hardware, or updated control logic---can alter the system dynamics so that previously observed behavior is no longer representative. A formal coverage guarantee after such a modification requires a sufficient number of post-change measurements; until these are available, no statistical guarantee can be provided. As an immediate and conservative safeguard (\textbf{Case~3} in \cref{fig:modelUpdates}), the previously identified uncertainty set is inflated to $\mathcal{W}'=\mathcal{W}+\Delta\mathcal{W}$ using a margin $\Delta\mathcal{W}$ derived from engineering judgment or prior experience.
Reachability analysis is then repeated to verify whether the enlarged reachable sets $\mathcal{Y}_{0:\nk-1}'$ still maintain a sufficient safety margin to the unsafe set. This yields a rapid, conservative assessment without a statistical coverage guarantee; once post-change measurements have been collected, $\mathcal{W}'$ is re-calibrated as in Case~1 or~2 to restore the guarantee.

\section{Composition of Reachset-Conformant Models} \label{subsec:composition}

Complex CPS often admit a decomposition into interacting subsystems that can be modeled and analyzed individually. Whether such subsystem-level analyses combine into a system-level guarantee is the central methodological question for applying reachset-conformant modeling in a development process.
We establish here that reachset conformance is preserved under
serial composition, where the output of subsystem $i-1$ serves as the input to subsystem $i$; parallel composition, where subsystems share a common input and their outputs are combined via a map $\imap_{\mathrm{y}}$; and feedback loops, where the input of subsystem $i$ additionally depends on its own past outputs via a map $\imap_{\mathrm{u}}$. In all three cases, the input and output signals may be transformed by interconnection maps $\imap$ (e.g., to model measurement or transmission errors), and the initial state of each subsystem is induced from the initial state of the interconnected system via a map $\imap_{\mathrm{x},i}$. The interconnection maps may be unknown, provided set-valued overapproximations $\imapA$ satisfying $\imap(v) \subseteq \imapA(v)$ for all $v$ are available---a mild requirement, as these follow from the interface specification, and $\imapA = \imap$ can be chosen whenever a map is known exactly. Throughout, the interconnected model is obtained by replacing every subsystem $S_{\mathrm{T},i}$ with its model $S_{\mathrm{M},i}$ and every interconnection map $\imap$ with its overapproximation $\imapA$. The proofs of all results in this section are provided in Appendix~\ref{app:proofs}.

\begin{figure}[t]
\begin{minipage}[t]{0.2\linewidth}
\centering
\begin{tikzpicture}[
    node distance=0.6cm,
    >=Latex,
    block/.style={
        draw, rounded corners, fill=white, thick,
        rectangle, minimum width=1cm,
        minimum height=0.6cm, align=center
    },
    circ/.style={
        draw, fill=white, thick,
        circle, inner sep=0,
        minimum size=0.8cm, align=center
    },
    line/.style={thick, ->}
]

\node (input) at (0,0) {};

\node[
    block,
    fill=mycolorLight,
    draw=mycolorDark,
    minimum width=2.5cm,
    minimum height=5.2cm,
    below=0.8cm of input,
    xshift=0.6cm,
    yshift=0.25cm
] (ST) {};

\node[block, below=0.9cm of input] (S1) {$S_{\mathrm{T},1}$};
\node[block, below=0.7cm of S1] (S2) {$S_{\mathrm{T},2}$};

\node[below=0.7cm of S2] (S3) {$\vdots$};

\node[block, below=0.7cm of S3] (SN) {$S_{\mathrm{T},N}$};

\node[above=0 of ST, text=mycolorDark, xshift=-0.9cm] {\huge $\sysT$ \qquad};

\node[below=0.7cm of SN] (output) {};

\draw[line]
(input) -- (S1) node[pos=0.4,right]
{$u_{0:k}(\sysT)$};
\draw[line]
(S1) -- (S2) node[pos=0.3,right]
{$y_{0:k}(S_{\mathrm{T},1})$};
\draw[line]
(S2) -- (S3) node[pos=0.3,right]
{$y_{0:k}(S_{\mathrm{T},2})$};
\draw[line]
(S3) -- (SN) node[pos=0.3,right]
{$y_{0:k}(S_{\mathrm{T},N-1})$};

\draw[line]
(SN) --
node[pos=0.75,right]
{$y_{0:k}(\sysT)$}
(output);

\end{tikzpicture}
\vspace{-0.4cm}
\caption{Serial composition $\sysT = S_{\mathrm{T},1} \circ \dots \circ S_{\mathrm{T},N}$.}\label{fig:serial}
\end{minipage} \hfill
\begin{minipage}[t]{0.56\linewidth}
\centering
\begin{tikzpicture}[
    node distance=0.6cm,
    >=Latex,
    block/.style={
        draw, rounded corners, fill=white, thick,
        rectangle, minimum width=1cm,
        minimum height=0.6cm, align=center
    },
    circ/.style={
        draw, fill=white, thick,
        circle, inner sep=0,
        minimum size=0.8cm, align=center
    },
    line/.style={thick, ->}
]

\node (input) at (0,0) {};

\node[
    block,
    fill=mycolorLight,
    draw=mycolorDark,
    minimum width=7cm,
    minimum height=3.4cm,
    below=0.8cm of input,
    xshift=0.2cm,
    yshift=0.2cm
] (ST) {};

\node[block, below=1.2cm of input, xshift=-2.6cm] (S1) {$S_{\mathrm{T},1}$};
\node[block, below=1.2cm of input, xshift=-0.7cm] (S2) {$S_{\mathrm{T},2}$};

\node[below=1.2cm of input, xshift=1cm] (S3) {};
\node[below=1.4cm of input, xshift=1cm] {$\dots$};

\node[block, below=1.2cm of input, xshift=2.2cm] (SN) {$S_{\mathrm{T},N}$};

\node[above=0 of ST, text=mycolorDark, xshift=-0.8cm] {\huge $\sysT$ \qquad};

\node[circ, below=3cm of input] (hy) {$\imap_{\mathrm{y}}$};
\node[below=0.7cm of hy] (output) {};

\draw[line]
(input) -- ++(0,-0.9) node[pos=0.5,right]
{$u_{0:k}(\sysT)$}
-|
(S2);

\draw[line] (input) ++(0,-0.9) -| (S1);
\draw[line] (input) ++(0,-0.9) -| (S3);
\draw[line] (input) ++(0,-0.9) -| (SN);

\draw[line]
(S1) -- ++(0,-1)
node[pos=0.5,right]
{$y_{0:k}(S_{\mathrm{T},1})$}
-- (hy.160);

\draw[line]
(S2) -- ++(0,-1)
node[pos=0.5,right]
{$y_{0:k}(S_{\mathrm{T},2})$}
-- (hy.120);

\draw[line]
(S3) ++(0,-0.8) -- ++(0,-0.4)
-- (hy.60);

\draw[line]
(SN) -- ++(0,-1)
node[pos=0.5,right]
{$y_{0:k}(S_{\mathrm{T},N})$}
-- (hy.20);

\draw[line]
(hy) --
node[pos=0.75,right]
{$y_{0:k}(\sysT)$}
(output);

\end{tikzpicture}
\vspace{-0.4cm}
\caption{Parallel composition $\sysT = S_{\mathrm{T},1} \| \cdots \| S_{\mathrm{T},N}$.}\label{fig:parallel}
\end{minipage} \hfill
\begin{minipage}[t]{0.2\linewidth}
\centering
\begin{tikzpicture}[
    node distance=0.6cm,
    >=Latex,
    block/.style={
        draw, rounded corners, fill=white, thick,
        rectangle, minimum width=1cm,
        minimum height=0.6cm, align=center
    },
    circ/.style={
        draw, fill=white, thick,
        circle, inner sep=0,
        minimum size=0.8cm, align=center
    },
    line/.style={thick, ->}
]

\node (input) at (0,0) {};

\node[
    block,
    fill=mycolorLight,
    draw=mycolorDark,
    minimum width=2.7cm,
    minimum height=2.9cm,
    below=0.8cm of input,
    yshift=0.2cm,
    xshift=0.1cm
] (ST) {};

\node[circ, below=0.9cm of input] (hu) {$\imap_{\mathrm{u}}$};

\node[block, below=0.8cm of hu] (S1) {$S_{\mathrm{T},i}$};

\node[below=0.9cm of S1] (output) {};

\node[above=0 of ST, text=mycolorDark, xshift=-1cm] {\huge $\sysT$ \qquad};

\draw[line] (input) -- (hu) node[pos=0.4,right] {$u_{0:k}(\sysT)$};

\draw[line] (hu) -- (S1) node[pos=0.45,right] {$u_{0:k}(S_{\mathrm{T},i})$};

\draw[line] (S1) -- (output)node[pos=0.7,right] {$y_{0:k}(\sysT)$};

\draw[line] (S1) ++ (0,-0.5) -- ++(-1,0)  |- (hu.west);

\end{tikzpicture}
\vspace{-0.4cm}
\caption{Feedback loop $\sysT = \circlearrowleft S_{\mathrm{T},i}$.}\label{fig:feedback}
\end{minipage}
\end{figure}

\begin{theorem}[Closure under Interconnection]\label{thm:composition}
Let the target system $\sysT$ be composed of subsystems $S_{\mathrm{T},i}$ via (a)~the serial composition $\sysT = S_{\mathrm{T},1} \circ \dots \circ S_{\mathrm{T},N}$ visualized in \cref{fig:serial}, (b)~the parallel composition $\sysT = S_{\mathrm{T},1} \| \cdots \| S_{\mathrm{T},N}$ visualized in \cref{fig:parallel}, or (c)~the feedback loop $\sysT = \circlearrowleft S_{\mathrm{T},i}$ around a single subsystem visualized in \cref{fig:feedback}, where the feedback is strictly causal, i.e., the input at time step $k$ depends only on the outputs up to time step $k-1$. We assume the availability of models $S_{\mathrm{M},i}$ that are reachset-conformant to $S_{\mathrm{T},i}$. Then the interconnected model $\sysM$
is reachset-conformant to $\sysT$.
\end{theorem}

Since the three interconnection operators may themselves be applied to interconnected systems, the closure property extends inductively to arbitrary architectures composed of these primitives.

\begin{corollary}[Arbitrary Interconnections]\label{cor:arbitrary}
Let $\sysT$ be any system obtained by recursively applying serial composition, parallel composition, and feedback interconnection to subsystems $S_{\mathrm{T},i}$, $i=1,\dots,N$, and let $S_{\mathrm{M},i}$ be reachset-conformant to $S_{\mathrm{T}, i}$ for all $i$. Then the interconnected model $\sysM$
is reachset-conformant to $\sysT$.
\end{corollary}

Corollary~\ref{cor:arbitrary} covers every architecture expressible as a block diagram over these primitives. The guarantee is compositional in the strict sense: no property of a subsystem model beyond its own reachset conformance is required, so subsystem models may be identified by different teams, from disjoint data sets, using different model structures, and at different points in the development timeline. We now show that the statistical coverage guarantee of \cref{prop:cover} composes as well, so that a system-level coverage bound follows from the subsystem-level guarantees without any system-level identification experiment.

\begin{theorem}[Composition of Coverage Guarantees]\label{thm:composedCov}
For each $i=1,\dots,N$, let reachset conformance of $S_{\mathrm{M},i}$ to $S_{\mathrm{T},i}$ be established by identifying $\nthetai$ parameters over $\nm$ test cases under \cref{ass:coverage}. Assume that, for each subsystem $i$, its identification test cases are exchangeable with the test case induced at subsystem $i$ by a new system-level test case $\tilde{m}$. Then the expected coverage probability of the interconnected model $\sysM$ satisfies \cref{eq:expectedCov} with $\ntheta=\sum_{i=1}^N \nthetai$.
\end{theorem}

The subsystem exchangeability assumption can, for example, be satisfied by requiring the subsystem test cases to be mutually consistent, which means the signals at their interfaces must coincide (e.g., the input of a subsystem in a serial composition must equal the output of the preceding subsystem), allowing the subsystem test cases to be composed into a system-level test case. The assumption then requires $\tilde{m}$ and the resulting system-level test cases, denoted by $\mathcal{M}$, to be exchangeable.

Notably, composition incurs no penalty in the coverage guarantee: identifying all $\ntheta=\sum_{i=1}^N \nthetai$ parameters monolithically over $\nm$ system-level test cases yields exactly the bound in \cref{eq:expectedCov}, yet the compositional route attains it by solving $N$ smaller problems on local measurements instead of one large problem on system-level measurements available only after integration. However, since the subsystem uncertainties are calibrated locally and the interconnections are overapproximated, the composed model is typically more conservative than a jointly calibrated monolithic one.

\section{Application to a Safety-Critical Brake System}\label{sec:brakeSys}

\begin{figure}
\centering
\input{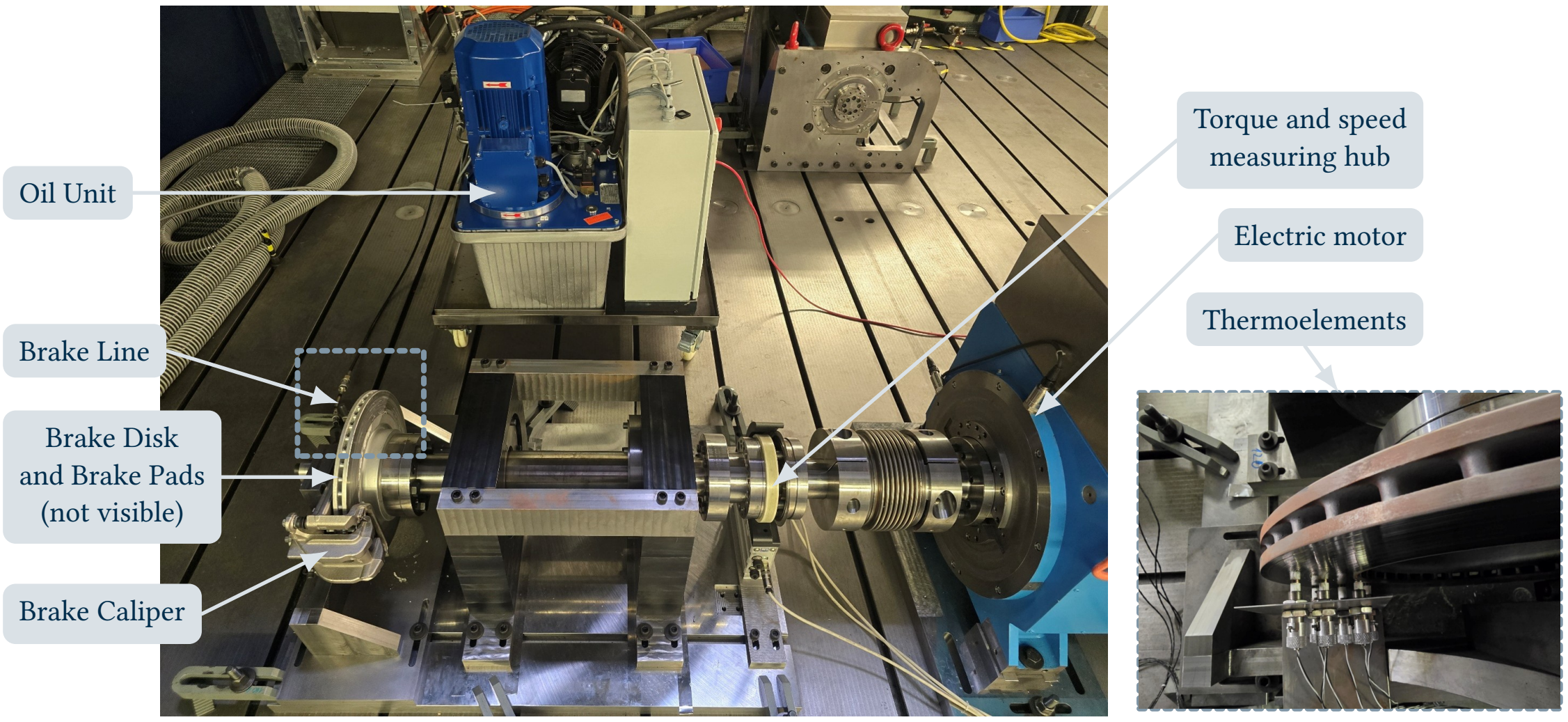}
\caption{Test bench of the brake system.}
\label{fig:brake}
\end{figure}

We now demonstrate the practical benefits outlined in \cref{subsec:benefits} on the automotive brake system shown in \cref{fig:brake}---a representative safety-critical CPS subject to stringent performance requirements and significant uncertainty from material properties, environmental conditions, and structural variability across subsystem designs. After introducing the system, its model, and the recorded data, the subsequent subsections address the four benefits: the unified treatment of uncertainty, the handling of late-stage modifications, the integration of black-box components, and compositional modeling.

\paragraph{System description} During operation, the hydraulic unit tracks a commanded reference pressure $p_{\mathrm{ref},k}$, while the caliper applies a normal force to the disk surface, generating a braking torque $\tau_k$ that depends on the caliper pressure $p_k$ and the instantaneous friction coefficient $\mu_k$. The friction coefficient varies substantially with material properties, temperature, run-in state, and surface contamination, making it a primary source of modeling uncertainty. Measurements $\hat{p}_k$ and $\hat{\tau}_k$ of the caliper pressure and braking torque, provided by the test bench, serve as input--output data for the modeling process. All experiments were performed on the same brake disk and pad without replacement, so that the recorded variability stems from the operating conditions rather than from different design choices. The disk has
an effective friction radius $\radius = 149.3$\,mm and
the caliper piston area is $\area = 5104$\,mm$^2$. The hydraulic unit exhibits a dead time of $0.14$\,s, corresponding to a delay of $\delay = 9$ time steps at the sampling time of $0.015$\,s.
The test-bench inertia is set to $98.0$\,kg\,m$^2$, representing the equivalent inertia assigned to one front brake disk of a battery electric vehicle, and a positive driving torque applied by the drive machine emulates the higher effective inertia of the vehicle.

\paragraph{Deterministic model} The relationships between reference pressure, caliper pressure, and braking torque are captured by the nonlinear state-space model in \cref{eq:generalSystem} with
\begin{subequations}\label{eq:brakeModel_determ}
\begin{align}
f(q,x_k,u_k,w_k) &=
\begin{bmatrix} x_{k(2)} \\ \vdots \\ x_{k(\delay)} \\
x_{k(\delay)} + \param_{\mathrm{u}(1)} (u_k - x_{k(1)}) + \param_{\mathrm{u}(2)} \\ x_{k({\delay}+1)} + \param_{\mathrm{u}(3)} \end{bmatrix}, && x_k = \begin{bmatrix}
    p_{k-{\delay+1}} \\ \vdots \\ p_{k-1} \\ p_k \\ \mu_k\end{bmatrix}, \qquad u_k = p_{\mathrm{ref},k-{\delay+1}}, \\
g(q,x_k,u_k,w_k) &=
\begin{bmatrix} x_{k(\delay)} + \param_{\mathrm{u}(4)} \\ \area \radius x_{k(\delay)} x_{k({\delay}+1)} + \param_{\mathrm{u}(5)} \end{bmatrix}, && y_k = \begin{bmatrix} \hat{p}_k \\ \hat{\tau}_k \end{bmatrix}, \label{eq:brakeModel_determ_y}
\end{align}
\end{subequations}
with $k\geq{\delay-1}$ and initial state $x_{{\delay-1}}=[\hat{p}_{0},~\dots,~ \hat{p}_{{\delay-1}},~ \mu_{{{\delay-1}},\mathrm{est}}+\param_{\mathrm{u}(6)}]^\top$, where the initial friction coefficient $\mu_{{{\delay-1}},\mathrm{est}}=\frac{1}{\delay} \sum_{i=0}^{{\delay-1}} \frac{\hat{\tau}_i}{\area\radius\hat{p}_{i}}$ is estimated from the initial pressure and torque measurements $\hat{p}_{i}$ and $\hat{\tau}_i$, $i=0,\dots,{\delay-1}$.
The parameters $\param_{\mathrm{u}} \in \mathbb{R}^6$ capture unmodeled effects on the pressure dynamics, the friction-coefficient evolution, and the measurement channels; since these effects depend on operating conditions and component realizations in ways not amenable to analytical modeling, $\param_{\mathrm{u}}$ is treated as unknown and identified from data.

\paragraph{Experimental procedure and data} Each of the eight brake test cases T1--T8 consists of a \emph{run-in} phase that establishes a defined frictional state, followed by a \emph{load stage} during which the measurements used below are recorded over 100 time steps at a sampling time of $0.015$\,s. Before every braking application, the test bench is accelerated to the specified initial speed and braking is initiated once the rotational speed has stabilized. Each run-in block comprises 20 braking applications, decelerating from an equivalent vehicle speed of $100$\,km/h to $50$\,km/h at a constant hydraulic pressure of $27$\,bar and subsequently to standstill at $19$\,bar, which conditions the friction pair under moderate load; the start of each application is governed by a test-case-specific temperature criterion, namely cooling below or preheating by slip operation above the run-in temperature. The subsequent load stage consists of a single braking application from $100$\,km/h at a constant reference pressure of $54$\,bar, imposing substantially higher normal force and contact pressure than the run-in. All test cases share this reference pressure sequence but differ in their operating conditions---the temperature and surface condition during run-in as well as the initial temperature and surface condition during operation---as summarized in \cref{tab:brake_experiments}: wet surface conditions are produced by water spray during run-in (T6) or during the load stage (T7), and the corroded surface in T8 results from water exposure followed by several days of storage before run-in. While these eight test cases suffice to demonstrate the modeling workflow, practically useful coverage guarantees would require considerably more test cases spanning a broader range of operating conditions and, in particular, varying input signals, since the guarantee of Prop.~\ref{prop:cover} extends only to future test cases exchangeable with the recorded ones.

\begin{table}[t]
\centering
\caption{Run-in and operation conditions for brake test cases T1--T8.}
\label{tab:brake_experiments}

\begin{tabular}{lrlrl}
\toprule
& \multicolumn{2}{c}{Run-in} & \multicolumn{2}{c}{Operation} \\
\cmidrule(lr){2-3} \cmidrule(lr){4-5}
Experiment & Temperature & Surface & Initial temperature & Surface \\
\midrule
T1 & $<200^{\circ}$C & dry & $[25,50)^{\circ}$C
& dry \\
T2 & $<400^{\circ}$C & dry & $[25,50)^{\circ}$C
& dry \\
T3 & $<400^{\circ}$C & dry & $[50,100)^{\circ}$C
& dry \\
T4 & $>200^{\circ}$C & dry & $[0,25)^{\circ}$C
& dry \\
T5 & $>400^{\circ}$C & dry & $[25,50)^{\circ}$C
& dry \\
T6 & $<200^{\circ}$C & wet & $[50,100)^{\circ}$C
& dry \\
T7 & $<200^{\circ}$C & dry & $[0,25)^{\circ}$C
& wet \\
T8 & $<200^{\circ}$C & dry \& corroded & $\geq100^{\circ}$C
& dry  \\
\bottomrule
\end{tabular}
\end{table}

\begin{figure}[t]
  \centering
  \setlength{\parindent}{0pt}%
  \sharedlegendBrake\par\vspace{0.2cm}%
  \begin{subfigure}[t]{\textwidth}
    \raggedright%
    \input{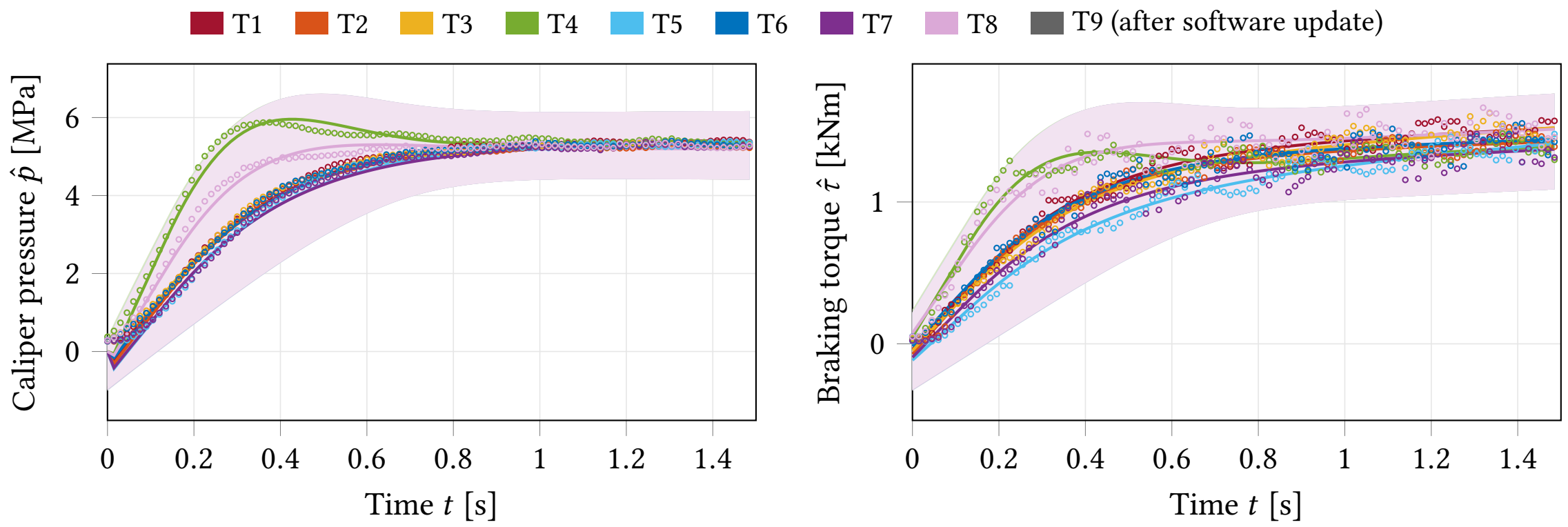}
    \caption{Structural model (\cref{subsec:captureUnc}): deterministic models fitted individually to each test case, and the single reachset-conformant model identified from T3--T8, whose reachable sets also enclose the unseen validation cases T1 and T2.}\label{fig:fig_example42}
  \end{subfigure}\par\vspace{0.3cm}%
  \begin{subfigure}[t]{0.49\textwidth}
    \centering
    \input{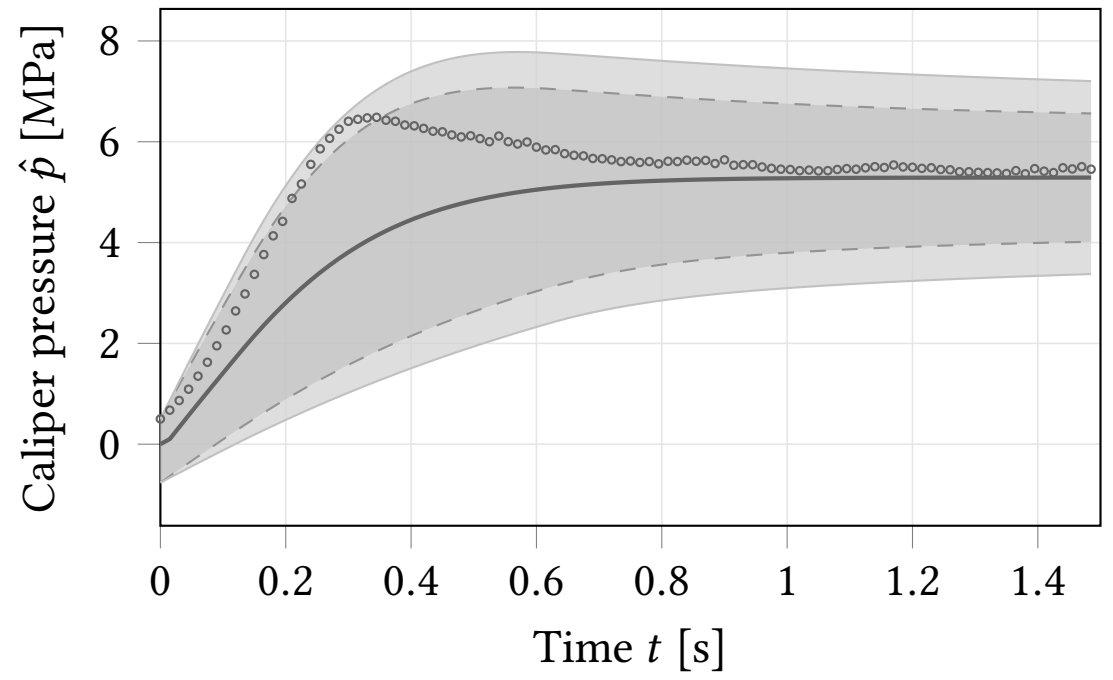}
    \caption{Software modification (\cref{subsec:lateStage}): prediction of the nominal pre-change model for the post-change test case T9, and reachable sets of the pre-change (dashed boundaries) and the recursively updated model (solid boundaries).}\label{fig:fig_example43}
  \end{subfigure}\hfill
  \begin{subfigure}[t]{0.49\textwidth}
    \centering
    \input{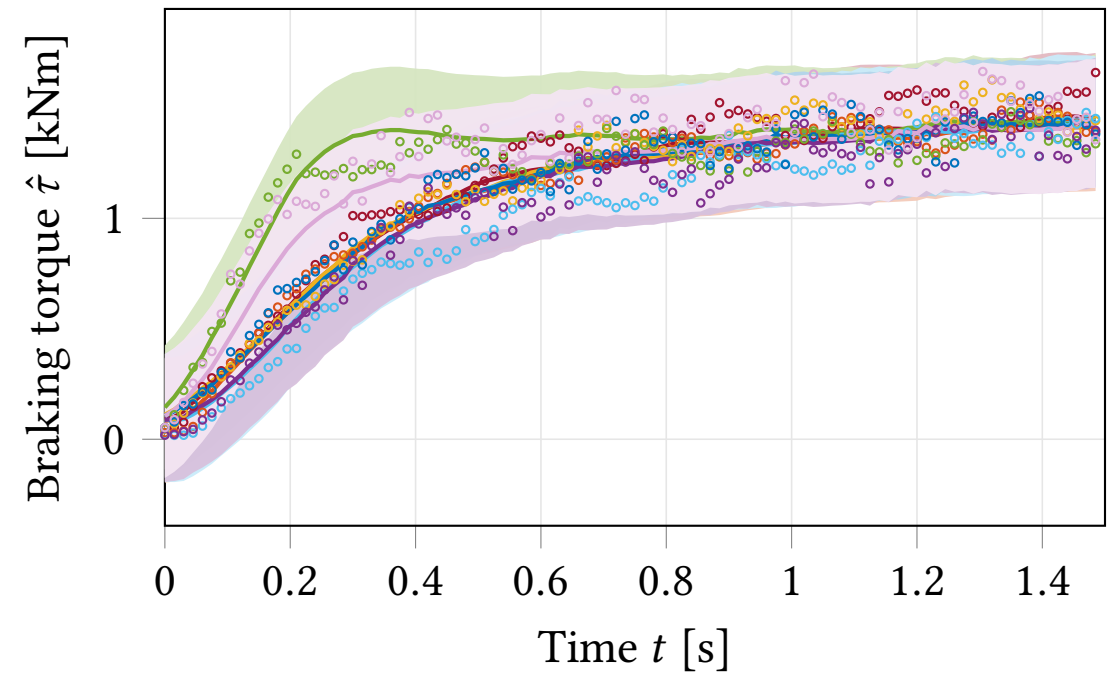}
    \caption{Structural friction model predicting the braking torque from measured caliper pressures (\cref{subsec:compBrake}): deterministic model fitted jointly over T1--T8 and reachset-conformant model $S_{\mathrm{M,\tau}}$ in \eqref{eq:torqueModel}.}\label{fig:fig_example45a}
  \end{subfigure}\par\vspace{0.3cm}%
  \begin{subfigure}[t]{0.49\textwidth}
    \centering
    \input{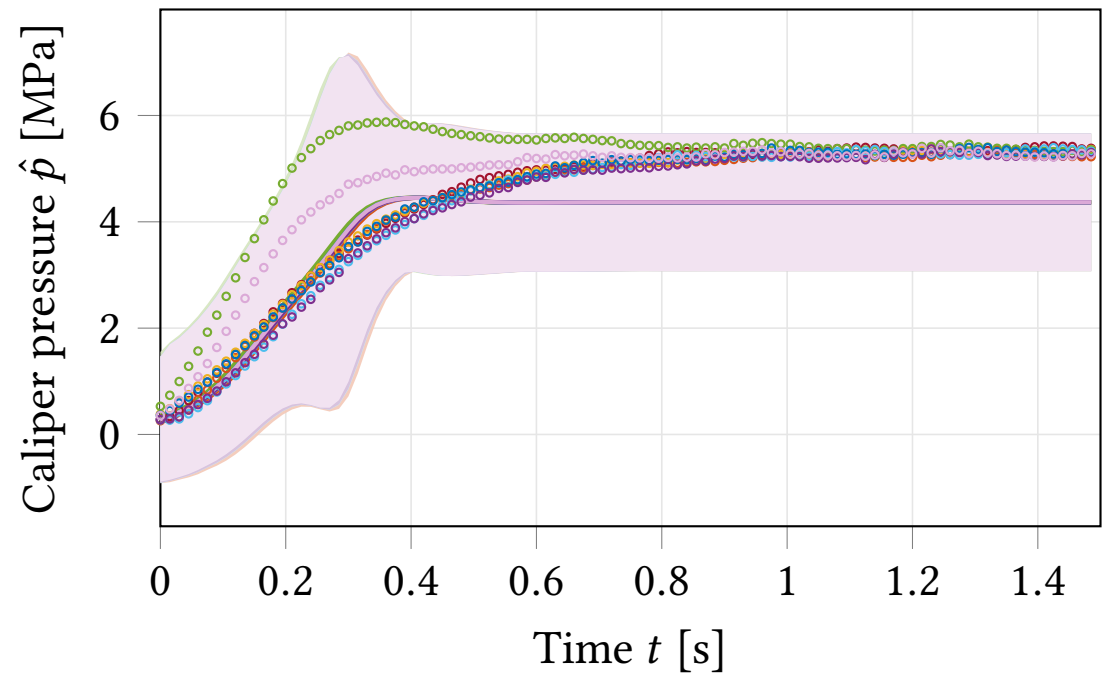}
    \caption{Black-box hydraulic unit (\cref{subsec:blackbox}): deterministic neural-network surrogate in \eqref{eq:pressureDynDet} trained on the pressure data of T1--T8 and reachset-conformant neural-network model $S_{\mathrm{M,p}}$ in \eqref{eq:pressureDyn}.}\label{fig:fig_example44}
  \end{subfigure}\hfill
  \begin{subfigure}[t]{0.49\textwidth}
    \centering
    \input{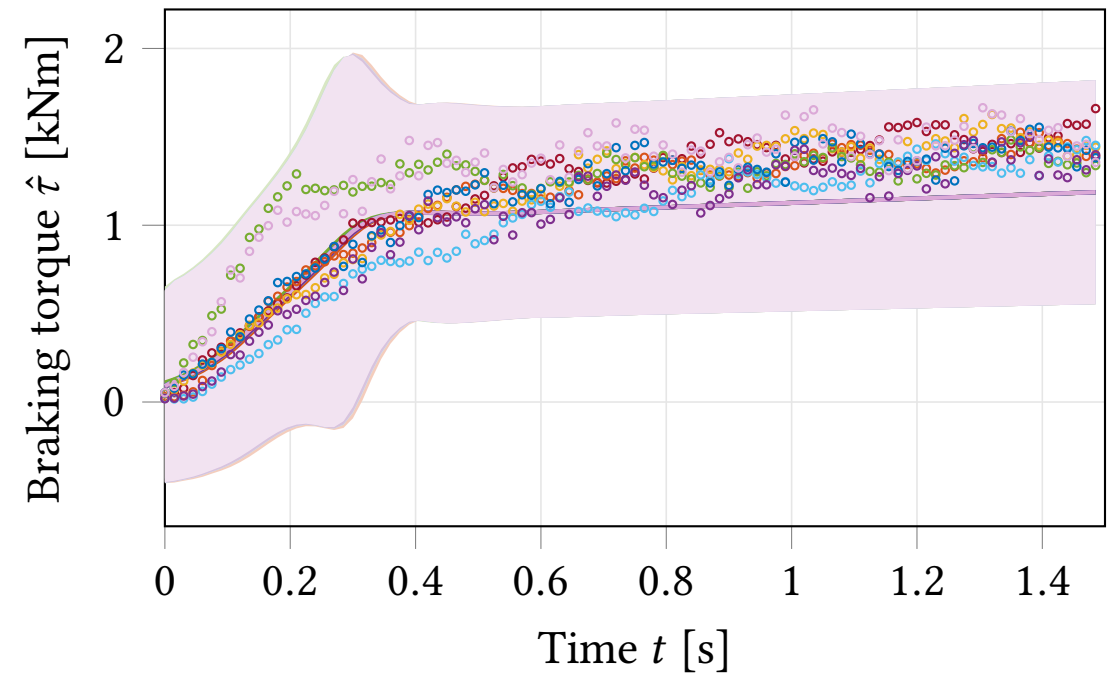}
    \caption{Composed model predicting the braking torque directly from the reference pressure (\cref{subsec:compBrake}): deterministic model fitted over T1--T8 and reachset-conformant model $S_{\mathrm{M,p}} \circ S_{\mathrm{M,\tau}}$ in \eqref{eq:composedModel}.}\label{fig:fig_example45b}
  \end{subfigure}
  \caption{Identification results for the brake system: measured trajectories (circular markers), predictions of the deterministic models (solid lines), and reachable sets of the reachset-conformant models (shaded areas).}
  \vspace{-1.1cm}
\end{figure}

\subsection{Unified Treatment of Uncertainty} \label{subsec:captureUnc}

In a conventional deterministic workflow, a separate instance of model \eqref{eq:brakeModel_determ} is identified for each relevant operating condition and hardware configuration: the parameters $\delay$, $\area$, and $\radius$ are fixed by the given hardware, while $\param_{\mathrm{u}}$ is identified from data recorded under the respective operating condition. Suppose the operating conditions of the brake system are discretized into the four run-in temperature conditions, three run-in surface conditions, four initial operation temperatures, and two operation surface conditions listed in \cref{tab:brake_experiments}. This coarse partition already yields $4 \cdot 3 \cdot 4 \cdot 2 = 96$ combinations, each requiring separate data collection, model identification, and validation. The choice of discretization granularity is itself nontrivial: a finer partition multiplies the identification
burden, while a coarser one risks large within-interval variation and poor predictive accuracy. Hardware variability further amplifies this effort: if the brake system must support $n_H$ hydraulic units with distinct delays $\delay$ and $n_C$ caliper variants with different piston areas $\area$, each hydraulic--caliper combination $(\delay,\area)$ defines a distinct nominal model whose residual parameters $\param_{\mathrm{u}}$ change through the modified pressure build-up and pad--disk interaction. The total number of identification and validation campaigns thus scales as $96 \cdot n_H \cdot n_C$, and every additional discretized factor---disk material, pad compound, or control-software version---multiplies this count further.
That this enumeration cannot simply be skipped when using deterministic models is evident from the data: \cref{tab:brake_identified_params} lists the parameters $\param_{\mathrm{u}}$ and the prediction errors of deterministic models fitted individually to each test case by minimizing the mean-square prediction error, alongside a single model fitted jointly over T3--T8. The individually fitted parameters deviate substantially across the test cases, while the jointly fitted model exhibits significantly larger prediction errors---no single deterministic model represents all operating conditions accurately.

\begin{table}[t]
\centering
\caption{Parameters $\param_{\mathrm{u}}$ and mean-square prediction errors $\mathrm{MSE}_{p}$ and $\mathrm{MSE}_{\tau}$ of deterministic models individually fitted to each test case and jointly fitted over T3--T8 (last row).}
\label{tab:brake_identified_params}
\begin{tabular}{lrrrrrrrr}
\toprule
Experiment & $\param_{\mathrm{u}(1)}$ & $\param_{\mathrm{u}(2)}$ & $\param_{\mathrm{u}(3)}$ & $\param_{\mathrm{u}(4)}$ & $\param_{\mathrm{u}(5)}$ & $\param_{\mathrm{u}(6)}$ & $\mathrm{MSE}_{p}$ & $\mathrm{MSE}_{\tau}$ \\
\midrule
T1 & 2.3424 & 2.52e-02 & 0.0005 & -0.70 & -0.1617 & 0.3164 & 1.03e-02 & 3.30e-03 \\
T2 & 2.3903 & 2.46e-02 & 0.0003 & -0.79 & -0.1876 & 0.3173 & 1.08e-02 & 2.59e-03 \\
T3 & 2.3345 & 1.87e-02 & 0.0007 & -0.57 & -0.1642 & 0.3068 & 1.09e-02 & 3.47e-03 \\
T4 & 4.0509 & 4.76e-02 & 0.0005 & -0.81 & -0.1065 & 0.2683 & 1.72e-02 & 4.90e-03 \\
T5 & 2.2182 & 3.10e-02 & 0.0008 & -0.91 & -0.2072 & 0.2558 & 2.27e-02 & 3.03e-03 \\
T6 & 2.3291 & 2.22e-02 & 0.0003 & -0.66 & -0.1377 & 0.3158 & 1.26e-02 & 4.52e-03 \\
T7 & 2.2436 & 2.84e-02 & 0.0004 & -0.90 & -0.2022 & 0.2924 & 2.08e-02 & 5.69e-03 \\
T8 & 3.2256 & 2.26e-02 & 0.0005 & -0.66 & -0.0567 & 0.3047 & 1.59e-02 & 8.27e-03 \\
\midrule
T3--T8 & 2.6624 & 2.08e-02 & 0.0005 & -0.63 & -0.1139 & 0.2880 & 1.75e-01 & 2.01e-02 \\
\bottomrule
\end{tabular}
\end{table}

Reachset-conformant identification avoids this enumeration entirely. We introduce uncertainty variables $w_k$ at the locations of $\param_{\mathrm{u}}$ in~\eqref{eq:brakeModel_determ}, i.e., we replace $\param_{\mathrm{u}(i)}$ with $(\param_{\mathrm{u}(i)}+w_{k(i)})$. Using the measurement trajectories of T3--T8---which jointly cover all individual operating conditions while leaving T1 and T2 as unseen validation cases---we first obtain $\param_{\mathrm{u}}$ by minimizing the mean-square prediction error (last row of~\cref{tab:brake_identified_params}) and then identify the uncertainty set
\begin{align*}
  \mathcal{W} = \left\{w \in \mathbb{R}^{6} \middle|~ |w| \leq \alpha\right\},
\end{align*}
where the bound $\alpha = [1.30, 0, 0, 0.63, 0.27, 0]^\top$ is obtained via linear programming \citep[Alg.~1]{luetzow2026reachset} with the identity matrix as generator template. \Cref{fig:fig_example42} compares the measured trajectories, the predictions of the individually fitted deterministic models, and the reachable sets of the reachset-conformant model, computed by linearization. The reachable sets are nearly identical across all test cases and enclose every measured trajectory, including those of the unseen validation cases T1 and T2. A single model thus suffices for all operating conditions, while unseen conditions are covered with a statistical guarantee.

Variations of the hardware parameters do not necessitate separate models either: instead of identifying one model per hydraulic--caliper combination, the affected parameters $\delay$ and $\area$ are replaced by bounded sets $\mathcal{Q}_{\mathrm{d}}$ and $\mathcal{Q}_{\mathrm{A}}$, and reachability analysis of a single model certifies all admissible combinations at once.

\subsection{Handling Late-Stage Modifications} \label{subsec:lateStage}

Late-stage design changes are incorporated through the model-update strategy from \cref{subsec:modelUpdates} rather than a re-identification from scratch, as we demonstrate for a controller update in which the proportional gain of the pressure controller is doubled and for which post-change measurements are available.
The delay $\delay$ remains unchanged, so no structural adaptation of the model equations in \cref{eq:brakeModel_determ} is required; the revised control logic instead alters the transient pressure build-up and relaxation behavior, which manifests as a change of the effective disturbance magnitudes captured by $\mathcal{W}$. To capture this change, a new test case T9 is recorded on the modified system. Since the target system has changed but post-change data is available, this situation corresponds to Case~2 in \cref{fig:fig_example43}: the uncertainty set is adapted with the recursive update procedure of \citet{luetzow2025recursive}, which enforces $\mathcal{W}' \supseteq \lambda \mathcal{W}$ via linear programming and requires only the new data set $\mathcal{M}'$ consisting of T9, without revisiting the original data set $\mathcal{M}$.

The forgetting factor is set to $\lambda = 0.99$ for two reasons. First, $\mathcal{M}'$ contains only a single test case, so aggressively discounting the past would effectively base the model on one trajectory and forfeit the robustness accumulated over T3--T8. Second, doubling the proportional gain is expected to modify the transient behavior but not to dramatically reduce the uncertainty. Choosing $\lambda = 0.99$ therefore retains most prior information while permitting gradual adaptation if the modified controller reduces the uncertainty; as further post-change test cases are collected, repeated recursive updates progressively shift the weight toward the behavior of the modified system.

\Cref{fig:fig_example43} shows the measured trajectory of T9 together with the reachable sets of the pre-update and the updated model. The recursive update enlarges the scaling factors from $\alpha = [1.30, 0, 0, 0.63, 0.27, 0]^\top$ (cf.~\cref{subsec:captureUnc}) to $\alpha' = [1.71, 0, 0, 0.63, 0.73, 0]^\top$. The faster pressure build-up induced by the doubled gain is not fully covered by the pre-update model between $t=0.2\,$s and $t=0.4\,$s, whereas the updated reachable sets $\mathcal{Y}_{0:\nk-1}'$ enclose the complete trajectory by construction. Repeating the reachability analysis with $\mathcal{W}'$ thus re-establishes the safety assessment for the modified system without restarting the identification workflow.

\subsection{Integrating Black-Box Components} \label{subsec:blackbox}

So far, the structure of the pressure dynamics in \cref{eq:brakeModel_determ} was derived from physical insight into the hydraulic unit. Assume now that the hydraulic unit is available only as a black box, providing reference pressures $p_{\mathrm{ref},k}$ and measured caliper pressures $\hat{p}_k$ as input--output data. Using the pressure data of T1--T8, a neural-network surrogate with 16 hidden neurons and tanh activation is trained to predict $\hat{p}_{k+1}$ from the past pressures $\hat{p}_{k-i}$ and reference pressures $p_{\mathrm{ref},k-i}$, $i=0,\dots,9$, assuming a hydraulic-unit delay of at most ten time steps:
\begin{align}\label{eq:pressureDynDet}
    \hat{p}_{k+1} = \param_{\mathrm{weights}2} \tanh\left( \param_{\mathrm{weights}1} \begin{bmatrix} p_{\mathrm{ref},k-9:k} \\ \hat{p}_{k-9:k} \end{bmatrix} + \param_{\mathrm{bias}1} \right) + \param_{\mathrm{bias}2}.
\end{align}
Due to approximation limits and finite training data, the network does not fit the true dynamics exactly; the resulting prediction errors, visible in \cref{fig:fig_example44}, render verification results untrustworthy in the traditional workflow.

To obtain a reachset-conformant model, bounded uncertainties $w_{\mathrm{p},k}$ are introduced at the network bias terms \citep{luetzow2025zonoconformal} and via a separate output equation, giving the nondeterministic pressure model $S_{\mathrm{M,p}}$:
\begin{subequations}\label{eq:pressureDyn}
\begin{align}
    p_{k+1} &= \param_{\mathrm{weights}2} \tanh\left( \param_{\mathrm{weights}1} \begin{bmatrix} p_{\mathrm{ref},k-9:k} \\ p_{k-9:k} \end{bmatrix} + \param_{\mathrm{bias}1} + w_{\mathrm{p},k(1:16)}\right) + \param_{\mathrm{bias}2} + w_{\mathrm{p},k(17)}, \\
    \hat{p}_k &= p_k + w_{\mathrm{p},k(18)},
\end{align}
\end{subequations}
with $k\geq 9$, state $p_{k-9:k}$, input $p_{\mathrm{ref},k-9:k}$, output $\hat{p}_k$, and initial state ${p}_{0:9}= \hat{p}_{0:9}$. The zero-centered uncertainty set
\begin{align*}    \mathcal{W}_{\mathrm{p}} = \left\{w \in \mathbb{R}^{18} \middle|~ |w| \leq \alpha_{\mathrm{p}}\right\}
\end{align*}
is calibrated such that all pressure measurements lie within the reachable set, yielding
\begin{align*}
    \alpha_{\mathrm{p}}=[0,0,0,0,0,0,0,0,1.74,0,0,0,0,0,0,0,0,0.34]^\top.
\end{align*}
As for the structural model in \cref{subsec:captureUnc}, the resulting reachable sets enclose all measured trajectories across all test cases and can be used directly for reachability-based verification; the wider reachable sets in \cref{fig:fig_example44} compared to \cref{fig:fig_example42} reflect the additional approximation uncertainty of the neural-network surrogate relative to the physics-based model structure in \cref{eq:brakeModel_determ}.

\subsection{Compositional Modeling} \label{subsec:compBrake}

Finally, we exploit the modular architecture of the brake system, which naturally decomposes into a hydraulic subsystem governing the pressure dynamics and a friction subsystem capturing the friction-coefficient evolution and its effect on the braking torque---a serial interconnection in the sense of \cref{subsec:composition}. This decomposition allows both subsystems to be identified independently: alongside the pressure neural-network model $S_{\mathrm{M,p}}$ from \cref{subsec:blackbox}, which is identified solely from reference and caliper pressure measurements, we identify a reachset-conformant friction model $S_{\mathrm{M,\tau}}$:
\begin{subequations}\label{eq:torqueModel}
\begin{align}
\mu_{k+1} &= \mu_{k} + \param_{\tau(1)} +  w_{\tau,k(1)}, \\
\hat{\tau}_k &= \area \radius \mu_{k} \hat{p}_k + \param_{\tau(2)} + w_{\tau,k(2)}, \label{eq:torqueModel_y}
\end{align}
\end{subequations}
with $k\geq 9$ (matching the input window of $S_{\mathrm{M,p}}$), state $\mu_k$, input $\hat{p}_k$, output $\hat{\tau}_k$, and initial state $\mu_{9} = \mu_{9,\mathrm{est}} + \param_{\tau(3)} + w_{\tau,9(3)}$, where $\mu_{9,\mathrm{est}}=\frac{1}{10} \sum_{i=0}^{9} \frac{\hat{\tau}_i}{\area\radius\hat{p}_{i}}$. This model predicts the braking torque $\hat{\tau}_k$ from given caliper pressure measurements $\hat{p}_k$. The parameters $\param_{\tau}=[0,0.03,0.29]^\top$ are identified by minimizing the mean-square prediction error over all test cases, and the uncertainty set
\begin{align*}
    \mathcal{W}_{\tau} = \left\{w \in \mathbb{R}^{3} \middle|~ |w| \leq \alpha_{\tau}\right\},
\end{align*}
with $\alpha_{\tau}=[0,0.28,0.001]^\top$, is calibrated by enforcing containment of all T1--T8 torque measurements within the reachable set of \cref{eq:torqueModel}; the resulting predictions are visualized in \cref{fig:fig_example45a}.

Connecting both subsystems by feeding the output of $S_{\mathrm{M,p}}$ as input to $S_{\mathrm{M,\tau}}$ yields the composed model $S_{\mathrm{M,p}} \circ S_{\mathrm{M,\tau}}$:
\begin{subequations}\label{eq:composedModel}
\begin{align}
\begin{bmatrix} p_{k+1} \\ \mu_{k+1} \end{bmatrix} &= \begin{bmatrix} \param_{\mathrm{weights}2} \tanh\left( \param_{\mathrm{weights}1} \begin{bmatrix} p_{\mathrm{ref},k-9:k} \\ p_{k-9:k} \end{bmatrix} + \param_{\mathrm{bias}1} + w_{\mathrm{p},k(1:16)}\right) + \param_{\mathrm{bias}2} + w_{\mathrm{p},k(17)} \\ \mu_{k} + \param_{\tau(1)} + w_{\tau,k(1)} \end{bmatrix}, \\
\hat{\tau}_k &= \area \radius \mu_{k} (p_k + w_{\mathrm{p},k(18)}) + \param_{\tau(2)} + w_{\tau,k(2)},
\end{align}
\end{subequations}
with $k\geq 9$ and initial state $[\hat{p}_{0},~\dots,~\hat{p}_{9},~ \mu_{9,\mathrm{est}}+\param_{\tau(3)}+w_{\tau,9(3)}]^\top$. This model predicts the braking torque directly from the reference pressure and a small number of initial caliper-pressure measurements, without requiring knowledge of the caliper-pressure evolution. Since both subsystem models are reachset-conformant and the interconnection map $u_k(S_{\mathrm{M,\tau}})= y_k(S_{\mathrm{M,p}})$ is exact, reachset conformance of the composed model follows directly from \cref{thm:composition}(a) without any additional identification. As shown in \cref{fig:fig_example45b}, the composed model $S_{\mathrm{M,p}} \circ S_{\mathrm{M,\tau}}$ is reachset-conformant but more conservative in predicting the braking torque than $S_{\mathrm{M,\tau}}$ alone (cf.~\cref{fig:fig_example45a}); when caliper-pressure measurements are available, $S_{\mathrm{M,\tau}}$ therefore remains the less conservative alternative. The composed model is also more conservative than the structural model from \cref{subsec:captureUnc} (cf.~\cref{fig:fig_example42}), since the latter exploits physical knowledge of the pressure dynamics and jointly calibrates the pressure and friction uncertainties to minimize the overall conservatism.

\section{Conclusions}\label{sec:concl}
This article presented reachset-conformant modeling as a practical framework for managing uncertainty in the development of cyber-physical systems. By predicting sets of behaviors that conservatively enclose the true system dynamics with statistical guarantees, reachset-conformant models expose unsafe behavior in early design stages, where corrections are inexpensive. Our main theoretical contribution establishes that reachset conformance is closed under serial composition, parallel composition, and feedback interconnection, so that independently identified subsystem models compose into a system-level model whose coverage guarantee follows from the subsystem guarantees. Complemented by update strategies that confine model adaptation after design changes to the affected subsystem, the framework was demonstrated on an automotive brake system subject to significant friction variability: a single model captured the uncertainty due to operating conditions and hardware variants, late-stage hardware and software modifications were absorbed without restarting the identification workflow, a black-box hydraulic unit was abstracted directly from input--output data, and the composition of independently identified pressure and friction models inherited reachset conformance by construction.

While the compositional guarantees come at the price of increased conservatism relative to jointly calibrated monolithic models, they enable verification to commence with the first component measurements and to proceed in parallel across subsystems. Future work will focus on reducing this conservatism, e.g., by exploiting correlations between subsystem uncertainties during composition, and on integrating the framework into established industrial development processes and toolchains, extending its applicability to larger interconnected architectures and further classes of safety-critical systems.

\bibliographystyle{abbrvnat}
\bibliography{literature,ipek_literature,literature2}

\appendix
\section{Uncertainty-Set Calibration via Linear Programming} \label{sec:lp}

In this section, we summarize the linear programming approach from \cite{luetzow2026reachset} for establishing reachset conformance.
We assume the model equations are given in the form of \cref{eq:generalSystem}. Then, we define $\mathcal{W}$ to be the zonotope (a special case of a centrally symmetric polytope)
    \begin{align*}
        \mathcal{W} = \left\{\sum_{i=1}^{\nu} \alpha_i \lambda_i G_{\mathrm{w}(\cdot,i)} \,\middle\vert\,\lambda_i\in[-1,1]\right\},
    \end{align*}
where the generator number $\nu$ and the generator matrix $G_{\mathrm{w}} \in \mathbb{R}^{n_{\mathrm{w}}\times \nu}$ are user-defined (e.g., we use $\nu=n_{\mathrm{w}}$ and $G_{\mathrm{w}}=\mathbf{I}$ in our experiments). The scaling factors are computed with the linear program
\begin{subequations} \label{eq:LP}
    \begin{align}
		\alpha^*,\beta^* =\argmin_{\alpha,\beta} &\sum_{m\in\mathcal{M}_{\mathrm{eval}}}\sum_{k=0}^{n_{\mathrm{k}}-1} w_k \mathbf{1}^\top  \sum_{i=0}^{k}|\bar{D}_{k,i}(x_0^{(m)},u_{0:k}^{(m)})G_{\mathrm{w}} | \alpha \hspace{-1.5cm}\label{eq:confSimpl_cost}\\
            \text{s.t. }
            \qquad  \mathbf{0} &\leq 	\alpha \label{eq:posconstraint},	 \\
  \begin{bmatrix}\mathbf{0} \\ \mathbf{0} \end{bmatrix} &\leq \begin{bmatrix}
            \mathrm{vert}_{k+1}\left( \mathbf{I}_{\nu}\right)\alpha + \beta^{(m)}_{k} \\
            \mathrm{vert}_{k+1}\left( \mathbf{I}_{\nu}\right) \alpha - \beta^{(m)}_{k} \\
        \end{bmatrix}, && \forall m\in\mathcal{M},\,k=0,...,n_{k}-1, \label{eq:conf_constr_gen1}\\
    y_{\Delta,k}^{(m)} &= \begin{bmatrix}
                 \bar{D}_{k,0}(x_0^{(m)},u_{0:k}^{(m)}) G_{\mathrm{w}}&\cdots&\bar{D}_{k,k}(x_0^{(m)},u_{0:k}^{(m)}) G_{\mathrm{w}}
            \end{bmatrix} \beta^{(m)}_{k}, && \forall m\in\mathcal{M},\,k=0,...,n_{k}-1,\label{eq:conf_constr_genEq}
	\end{align}
\end{subequations}
where $y_{\Delta,k}^{(m)} = y_{k}^{(m)}(S_T)-  \bar{y}_k(x_0^{(m)},u_{0:k}^{(m)})$, the vector $\beta$ consists of the additional optimization variables $\beta_k^{(m)}\in\mathbb{R}^{(k+1)\nu}$, and  $\mathrm{vert}_{k+1}\left( \mathbf{I}_{\nu}\right)$ is the vertical concatenation of $k+1$ identity matrices of dimensions $\nu\times \nu$. The nominal output $\bar{y}_k(x_0,u_{0:k})$ and the linearized disturbance matrix $\bar{D}_{k,i}(x_0,u_{0:k})$ are computed as
\begin{subequations}
    \label{eq:ss_D}
    \begin{align*}
    \bar{y}_k(x_0,u_{0:k}) &= g(\bar{x}_k,u_k,\mathbf{0}), \\
    \bar{D}_{k,i}(x_0,u_{0:k})  &= \begin{cases}
            C_{k}(x_0,u_{0:k})\left(\prod_{l=1}^{k-i-1} A_{k-l}(x_0,u_{0:k})\right) B_{i}(x_0,u_{0:k}) & 0\leq i < k, \\
            D_{k}(x_0,u_{0:k}) & i=k,
            \end{cases} \\
        A_{k}(x_0,u_{0:k}) &= \nabla_{{x}_k}~f(q,x_{k},u_k,w_k)|_{x_k=\bar{x}_{k},w_k=\mathbf{0}}, \qquad
		B_{k}(x_0,u_{0:k}) =  \nabla_{{w}_k}~f(q,{x}_{k},u_k,w_k)|_{x_k=\bar{x}_{k},w_k=\mathbf{0}}, \\
		C_{k}(x_0,u_{0:k}) &= \nabla_{{x}_k}~g(q,x_{k},u_k,w_k)|_{x_k=\bar{x}_{k},w_k=\mathbf{0}}, \qquad
		D_{k}(x_0,u_{0:k}) =  \nabla_{{w}_k}~g(q,{x}_{k},u_k,w_k)|_{x_k=\bar{x}_{k},w_k=\mathbf{0}},
    \end{align*}
\end{subequations}
where the predicted state $\bar{x}_k$ is obtained from recursively evaluating \cref{eq:generalSystemState} starting at the initial state $\bar{x}_0=x_0$ and applying the inputs $u_{0:k}$ while setting all uncertainties $w_{0:k}$ to $\mathbf{0}$.

By using $G_{\mathrm{w}}=\mathbf{I}$, the resulting uncertainty set reduces to a multi-dimensional interval with half-widths $\alpha$, i.e.,
  $\mathcal{W} = \left\{w \in \mathbb{R}^{n_{\mathrm{w}}} \middle|~ |w| \leq \alpha\right\}$,
while more flexible shapes are possible using non-diagonal generator matrices.

\section{Proofs}\label{app:proofs}
This appendix collects the proofs for the coverage and the compositional guarantees, following the order in which the results appear in the main text.

\subsection{Coverage Guarantees}
We begin by establishing the coverage result stated in \cref{sec:rcModeling}.

\begin{proof}[Proof of Prop.~\ref{prop:cover}]
We first address the static setting, then lift the result to dynamical systems via output stacking.

\textit{Static case.} Let $\bar{\mathcal{Y}}^{(\tilde{m})}(\sysM)$ be a prediction set, $\bar{y}^{(\tilde{m})}(\sysT)$ a measurement, and $\bar{\eta}(\mathcal{M}) = \mathbb{P}\{\bar{y}^{(\tilde{m})}(\sysT) \in \bar{\mathcal{Y}}^{(\tilde{m})}(\sysM)\}$ the coverage probability. The bound in \cref{eq:expectedCov} then follows from \citet[Eq.~(9)]{luetzow2025zonoconformal} with $n_{\mathrm{out}} = 0$, since no outliers are admitted during identification; see \citet[Sec.~5.2]{luetzow2025zonoconformal} for details.

\textit{Dynamical case.} Define the stacked reachable set and stacked measurement over the horizon $n_k$ as
\begin{equation*}
\bar{\mathcal{Y}}^{(m)}(\sysM) := \mathcal{Y}_{0:n_k-1}^{(m)}(\sysM), \qquad \bar{y}^{(m)}(\sysT) := y_{0:n_k-1}^{(m)}(\sysT).
\end{equation*}
This recasts the dynamical system as a static mapping from initial states and input sequences to stacked outputs. The coverage probability then coincides with the definition given in \cref{sec:rcModeling}, and the result follows by direct application of the static case.
\end{proof}

\subsection{Composition of Subsystems}

Throughout this section, we use the interconnection maps $\imap$ and their overapproximations $\imapA$ introduced in \cref{subsec:composition}, where the maps may be non-deterministic but are assumed to be bounded. Since the interconnected model must satisfy \cref{eq:reachConf} for every admissible initial state and input sequence, we fix an arbitrary admissible pair throughout and denote it by $x_{0}^{(m)}(\sysT)$ and $u_{0:k}^{(m)}(\sysT)$, in analogy to the test-case notation of \cref{sec:rcModeling}. The initial states and inputs that this pair can induce for subsystem $S_{\mathrm{T},i}$ are collected in the sets $\mathcal{X}_0^{(m)}(S_{\mathrm{T},i})$ and $\mathcal{U}_{0:k}^{(m)}(S_{\mathrm{T},i})$, respectively, and the corresponding reachable output sets are denoted by
\begin{align*}
    \mathcal{Y}_k^{(m)}(S_{\mathrm{T},i}):=\mathcal{Y}_k(S_{\mathrm{T},i},\mathcal{X}_0^{(m)}(S_{\mathrm{T},i}),\mathcal{U}_{0:k}^{(m)}(S_{\mathrm{T},i})),
\end{align*}
with analogous definitions for the interconnected model $\sysM$.

The proofs share a common two-step structure: we first establish containment of the subsystem-output sets and then lift this containment to the interconnected system. We factor the recurring arguments into three lemmas.

\begin{lemma}[Containment transfer]\label{lem:transfer}
Let $\imap$ be an interconnection map with overapproximation $\imapA$, and let $\mathcal{V}_j \subseteq \mathcal{W}_j$ for $j=1,\dots,n$. Then $\imap(\mathcal{V}_1,\dots,\mathcal{V}_n) \subseteq \imapA(\mathcal{W}_1,\dots,\mathcal{W}_n)$.
\end{lemma}

\begin{proof}
Elementwise evaluation yields $\imap(\mathcal{V}_1,\dots,\mathcal{V}_n) \subseteq \imapA(\mathcal{V}_1,\dots,\mathcal{V}_n) \subseteq \imapA(\mathcal{W}_1,\dots,\mathcal{W}_n)$, where the first inclusion holds since $\imapA$ overapproximates $\imap$, and the second since $\mathcal{V}_j \subseteq \mathcal{W}_j$ for $j=1,\dots,n$.
\end{proof}

The initial states are interconnected identically in all three compositions, i.e., $x_{0}(S_{\mathrm{T},i}) \in \imap_{\mathrm{x},i}(x_{0}(\sysT))$, so that $\mathcal{X}_0^{(m)}(S_{\mathrm{T},i}) = \imap_{\mathrm{x},i}(x_{0}^{(m)}(\sysT))$ and $\mathcal{X}_0^{(m)}(S_{\mathrm{M},i}) = \imapA_{\mathrm{x},i}(x_{0}^{(m)}(\sysT))$; Lemma~\ref{lem:transfer} thus immediately yields
\begin{equation}\label{eq:x0Containment}
\mathcal{X}_0^{(m)}(S_{\mathrm{T},i}) \subseteq \mathcal{X}_0^{(m)}(S_{\mathrm{M},i}), \qquad \forall i.
\end{equation}
The next lemma establishes the first step, containment at the subsystem level; by \cref{eq:x0Containment}, only the input containment remains to be verified in each proof.

\begin{lemma}[Monotonicity]\label{lem:monotone}
Let $S_{\mathrm{M},i}$ be reachset-conformant to $S_{\mathrm{T},i}$ according to \cref{eq:reachConf}, and let $\mathcal{U}_{0:j}^{(m)}(S_{\mathrm{T},i}) \subseteq \mathcal{U}_{0:j}^{(m)}(S_{\mathrm{M},i})$ for all $j\leq k$. Then
\begin{equation}\label{eq:monotone}
\mathcal{Y}_{0:k}^{(m)}(S_{\mathrm{T},i}) \subseteq \mathcal{Y}_{0:k}^{(m)}(S_{\mathrm{M},i}).
\end{equation}
\end{lemma}

\begin{proof}
Fix $j\leq k$ and consider any $x_0\in\mathcal{X}_0^{(m)}(S_{\mathrm{T},i})$ and $u_{0:j}\in\mathcal{U}_{0:j}^{(m)}(S_{\mathrm{T},i})$. Reachset conformance of $S_{\mathrm{M},i}$ yields $\mathcal{Y}_j(S_{\mathrm{T},i},x_0,u_{0:j}) \subseteq \mathcal{Y}_j(S_{\mathrm{M},i},x_0,u_{0:j})$.
Taking the union over all such pairs gives $\mathcal{Y}_j^{(m)}(S_{\mathrm{T},i}) \subseteq \mathcal{Y}_j(S_{\mathrm{M},i}, \mathcal{X}_0^{(m)}(S_{\mathrm{T},i}),\mathcal{U}_{0:k}^{(m)}(S_{\mathrm{T},i})) \subseteq \mathcal{Y}_j^{(m)}(S_{\mathrm{M},i})$, where the second inclusion uses \cref{eq:x0Containment} and the assumed input containment. Because the stacked sets are Cartesian products of the time-step-wise sets, and Cartesian products preserve set inclusion, \cref{eq:monotone} follows.
\end{proof}

The second step, lifting the subsystem-level containment to the interconnected system, is captured by the following lemma.

\begin{lemma}[Output lifting]\label{lem:lift}
Let the outputs of the interconnected system and model be given by $y_{0:k}(\sysT) \in \imap_{\mathrm{y}}(y_{0:k}(S_{\mathrm{T},j_1}),\dots,y_{0:k}(S_{\mathrm{T},j_n}))$ and $y_{0:k}(\sysM) \in \imapA_{\mathrm{y}}(y_{0:k}(S_{\mathrm{M},j_1}),\dots,y_{0:k}(S_{\mathrm{M},j_n}))$. If $\mathcal{Y}_{0:k}^{(m)}(S_{\mathrm{T},j}) \subseteq \mathcal{Y}_{0:k}^{(m)}(S_{\mathrm{M},j})$ for all $k$, $m$ and all $j\in\{j_1,\dots,j_n\}$, then $\sysM$ is reachset-conformant to $\sysT$ according to \cref{eq:reachConf}.
\end{lemma}

\begin{proof}
Applying Lemma~\ref{lem:transfer} to the output interconnection yields
\begin{align*}
\mathcal{Y}_{0:k}^{(m)}(\sysT) = \imap_{\mathrm{y}}(\mathcal{Y}_{0:k}^{(m)}(S_{\mathrm{T},j_1}),\dots,\mathcal{Y}_{0:k}^{(m)}(S_{\mathrm{T},j_n})) \subseteq \imapA_{\mathrm{y}}(\mathcal{Y}_{0:k}^{(m)}(S_{\mathrm{M},j_1}),\dots,\mathcal{Y}_{0:k}^{(m)}(S_{\mathrm{M},j_n})) = \mathcal{Y}_{0:k}^{(m)}(\sysM).
\end{align*}
Projecting this inclusion onto the components of time step $k$ shows that $\mathcal{Y}_k^{(m)}(\sysT) \subseteq \mathcal{Y}_k^{(m)}(\sysM)$ for all $k$ and $m$. Thus, reachset conformance of $\sysM$ to $\sysT$ according to \cref{eq:reachConf} follows.
\end{proof}

\subsubsection{Reachset Conformance of Composed Subsystem Models}
\begin{proof}[Proof of \cref{thm:composition}(a): Serial Composition]
The inputs and outputs of $\sysT = S_{\mathrm{T},1} \circ \dots \circ S_{\mathrm{T},N}$ are interconnected via
\begin{align*}
    u_{0:k}(S_{\mathrm{T},i}) &\in \begin{cases}\imap_{\mathrm{u},1}(u_{0:k}(\sysT)) & \text{if }i=1, \\
    \imap_{\mathrm{u},i}(y_{0:k}(S_{\mathrm{T},i-1})) & \text{else}, \end{cases}\\
    y_{0:k}(\sysT) &\in \imap_{\mathrm{y}}(y_{0:k}(S_{\mathrm{T},N})),
\end{align*}
so that $\mathcal{U}_{0:k}^{(m)}(S_{\mathrm{T},1}) = \imap_{\mathrm{u},1}(u_{0:k}^{(m)}(\sysT))$ and $\mathcal{U}_{0:k}^{(m)}(S_{\mathrm{T},i}) = \imap_{\mathrm{u},i}(\mathcal{Y}_{0:k}^{(m)}(S_{\mathrm{T},i-1}))$ for $i\geq 2$. We show by induction on $i$ that
\begin{equation}\label{eq:claimSerial}
\mathcal{Y}_{0:k}^{(m)}(S_{\mathrm{T},i}) \subseteq \mathcal{Y}_{0:k}^{(m)}(S_{\mathrm{M},i}), \qquad \forall k.
\end{equation}
Applying Lemma~\ref{lem:transfer} to the input interconnection---with argument $u_{0:k}^{(m)}(\sysT)$ for $i=1$, and with $\mathcal{Y}_{0:k}^{(m)}(S_{\mathrm{T},i-1}) \subseteq \mathcal{Y}_{0:k}^{(m)}(S_{\mathrm{M},i-1})$ from the induction hypothesis for $i\geq 2$---yields $\mathcal{U}_{0:k}^{(m)}(S_{\mathrm{T},i}) \subseteq \mathcal{U}_{0:k}^{(m)}(S_{\mathrm{M},i})$ for all $k$, and Lemma~\ref{lem:monotone} establishes \cref{eq:claimSerial} for $i$, completing the induction. Finally, applying Lemma~\ref{lem:lift} to \cref{eq:claimSerial} for $i=N$ yields reachset conformance of $\sysM$ to $\sysT$.
\end{proof}

\begin{proof}[Proof of \cref{thm:composition}(b): Parallel Composition]
The inputs and outputs of $\sysT = S_{\mathrm{T},1} \| \dots \| S_{\mathrm{T},N}$ are interconnected via
\begin{align*}
    u_{0:k}(S_{\mathrm{T},i}) &\in \imap_{\mathrm{u},i}(u_{0:k}(\sysT)), \\
    y_{0:k}(\sysT) &\in \imap_{\mathrm{y}}(y_{0:k}(S_{\mathrm{T},1}),\dots,y_{0:k}(S_{\mathrm{T},N})),
\end{align*}
so that $\mathcal{U}_{0:k}^{(m)}(S_{\mathrm{T},i}) = \imap_{\mathrm{u},i}(u_{0:k}^{(m)}(\sysT))$. For each $i$, applying Lemma~\ref{lem:transfer} to the input interconnection yields $\mathcal{U}_{0:k}^{(m)}(S_{\mathrm{T},i}) \subseteq \mathcal{U}_{0:k}^{(m)}(S_{\mathrm{M},i})$ for all $k$, so that Lemma~\ref{lem:monotone} gives $\mathcal{Y}_{0:k}^{(m)}(S_{\mathrm{T},i}) \subseteq \mathcal{Y}_{0:k}^{(m)}(S_{\mathrm{M},i})$ for all $i$ and $k$, and Lemma~\ref{lem:lift} yields reachset conformance of $\sysM$ to $\sysT$. Note that evaluating $\imap_{\mathrm{y}}$ on all tuples from the Cartesian product of the subsystem output sets, as in Lemma~\ref{lem:lift}, ignores the coupling induced by the shared input; this is a sound overapproximation since the tuple of realized trajectories is always contained in this product.
\end{proof}

\begin{proof}[Proof of \cref{thm:composition}(c): Feedback Loop]
The inputs and outputs of $\sysT = \circlearrowleft S_{\mathrm{T},i}$ are interconnected via
\begin{align*}
    u_{0}(S_{\mathrm{T},i}) &\in \imap_{\mathrm{u},i}(u_{0}(\sysT)), ~\text{ and }~
    u_{0:k}(S_{\mathrm{T},i}) \in \imap_{\mathrm{u},i}(u_{0:k}(\sysT),y_{0:k-1}(S_{\mathrm{T},i})), ~k\geq1,\\
    y_{0:k}(\sysT) &\in \imap_{\mathrm{y}}(y_{0:k}(S_{\mathrm{T},i})),
\end{align*}
reflecting the strict causality of the feedback assumed in \cref{thm:composition}. Accordingly, the input sets are defined recursively over time by $\mathcal{U}_{0}^{(m)}(S_{\mathrm{T},i}) = \imap_{\mathrm{u},i}(u_{0}^{(m)}(\sysT))$ and $\mathcal{U}_{0:k}^{(m)}(S_{\mathrm{T},i}) = \imap_{\mathrm{u},i}(u_{0:k}^{(m)}(\sysT),\mathcal{Y}_{0:k-1}^{(m)}(S_{\mathrm{T},i}))$ for $k\geq 1$. We show by induction on the time step $k$ that
\begin{equation}\label{eq:claimFeedback}
\mathcal{U}_{0:k}^{(m)}(S_{\mathrm{T},i}) \subseteq \mathcal{U}_{0:k}^{(m)}(S_{\mathrm{M},i}) \quad\text{and}\quad \mathcal{Y}_{0:k}^{(m)}(S_{\mathrm{T},i}) \subseteq \mathcal{Y}_{0:k}^{(m)}(S_{\mathrm{M},i}).
\end{equation}
Applying Lemma~\ref{lem:transfer} to the input interconnection---with argument $u_{0}^{(m)}(\sysT)$ for $k=0$, and with $\mathcal{Y}_{0:k-1}^{(m)}(S_{\mathrm{T},i}) \subseteq \mathcal{Y}_{0:k-1}^{(m)}(S_{\mathrm{M},i})$ from the induction hypothesis for $k\geq 1$---yields the input containment in \cref{eq:claimFeedback}. Since the input containment holds up to horizon $k$, Lemma~\ref{lem:monotone} establishes the output containment, completing the induction. Finally, Lemma~\ref{lem:lift} yields reachset conformance of $\sysM$ to $\sysT$.
\end{proof}

\begin{proof}[Proof of Corollary~\ref{cor:arbitrary}]
Structural induction over the interconnection term. The base case is a single subsystem, for which the claim holds by assumption. For the induction step, each operator preserves reachset conformance by \cref{thm:composition}, applied to the operands, which are reachset-conformant by the induction hypothesis.
\end{proof}

\subsubsection{Coverage Guarantee of Composed Subsystem Models}
\begin{proof}[Proof of \cref{thm:composedCov}]
We condition on the identified models by fixing the training data $\mathcal{M}$ and consider the new system-level test case $\tilde{m}$. Since $\tilde{m}$ is a genuine experiment on the interconnected system, its realized trajectory is consistent at every interface, so each subsystem $S_{\mathrm{T},i}$ is driven by its in-situ input $u_{0:\nk-1}^{(\tilde{m})}(S_{\mathrm{T},i})$. Let
\begin{align*}
    C_i := \bigl\{\, y_{0:\nk-1}^{(\tilde{m})}(S_{\mathrm{T},i}) \in \mathcal{Y}_{0:\nk-1}\bigl(S_{\mathrm{M},i},\, x_0^{(\tilde{m})}(S_{\mathrm{T},i}),\, u_{0:\nk-1}^{(\tilde{m})}(S_{\mathrm{T},i})\bigr) \,\bigr\}
\end{align*}
denote the event that subsystem $i$ covers $\tilde{m}$, i.e., its measured output lies in the reachable set of $S_{\mathrm{M},i}$ evaluated at its measured initial state and input; by definition, $\mathbb{P}(C_i \mid \mathcal{M}) = \eta_i(\mathcal{M})$ is the per-subsystem coverage probability of \cref{eq:coverageProb} applied to subsystem $i$. Correspondingly, $E := \{\, y_{0:\nk-1}^{(\tilde{m})}(\sysT) \in \mathcal{Y}_{0:\nk-1}^{(\tilde{m})}(\sysM) \,\}$ is the system-level coverage event with $\mathbb{P}(E \mid \mathcal{M}) = \eta(\mathcal{M})$.

We first show that joint subsystem coverage implies system coverage, i.e.,
\begin{align}\label{eq:pathwiseComp}
    \textstyle\bigcap_{i=1}^N C_i \subseteq E.
\end{align}
This is the pathwise specialization of the proof of \cref{thm:composition}, obtained by replacing the measured-trajectory sets $\mathcal{Y}_{0:\nk-1}^{(m)}(\cdot)$ by the single realized trajectory of $\tilde{m}$: whenever an upstream subsystem covers $\tilde{m}$, the measured subsystem output is enclosed by the corresponding model reachable set. Thus, containment transfer (Lemma~\ref{lem:transfer}) places the induced downstream input in the propagated model input set, and the input-monotonicity of the reachable sets (Lemma~\ref{lem:monotone}) then encloses the next measured output; iterating along the interconnection and lifting through the output map (Lemma~\ref{lem:lift}) encloses the measured system output, which establishes \cref{eq:pathwiseComp}. Taking complements, a system miss forces at least one subsystem miss, $\neg E \subseteq \bigcup_{i=1}^N \neg C_i$.

Conditioning on $\mathcal{M}$ and applying the union bound to this inclusion gives
\begin{align*}
    1 - \eta(\mathcal{M}) = \mathbb{P}(\neg E \mid \mathcal{M}) \leq \sum_{i=1}^N \mathbb{P}(\neg C_i \mid \mathcal{M}) = \sum_{i=1}^N \bigl(1 - \eta_i(\mathcal{M})\bigr),
\end{align*}
where no assumption on the joint distribution of the subsystem miss events is needed, as the union bound holds for arbitrary dependence. Taking the expectation over $\mathcal{M}$ and using linearity yields $\mathbb{E}\{\eta(\mathcal{M})\} \geq 1 - \sum_{i=1}^N \bigl(1 - \mathbb{E}\{\eta_i(\mathcal{M})\}\bigr)$. By assumption, the identification test cases of subsystem $i$ are exchangeable with the test case induced by $\tilde{m}$, so Prop.~\ref{prop:cover} applies to each subsystem and gives $\mathbb{E}\{\eta_i(\mathcal{M})\} \geq 1 - \nthetai/(\nm+1)$. Substituting,
\begin{align*}
    \mathbb{E}\{\eta(\mathcal{M})\} \geq 1 - \sum_{i=1}^N \frac{\nthetai}{\nm+1} = 1 - \frac{\ntheta}{\nm+1},
\end{align*}
which equals \cref{eq:expectedCov} with $\ntheta = \sum_{i=1}^N \nthetai$.
\end{proof}

\end{document}